\documentclass[11pt]{article}
\pdfoutput=1

\usepackage{jheppub}
\usepackage{url,comment}
\usepackage{times}
\usepackage{latexsym}
\usepackage{graphicx, graphics, hyperref, amsmath, amssymb, slashed, color, bbm,amsthm, array}
\usepackage[usenames,dvipsnames]{xcolor}
 \usepackage{subfigure}
\usepackage{mdwlist}
\usepackage{multirow}
\usepackage[normalem]{ulem}
\usepackage[e]{esvect}
\usepackage{cancel}
\usepackage{pstricks}
\usepackage[toc,page]{appendix}
\usepackage{color}
\usepackage{enumitem}

 \DeclareMathOperator{\ev}{eV} \DeclareMathOperator{\kev}{keV} \DeclareMathOperator{\mev}{MeV} \DeclareMathOperator{\gev}{GeV}  \DeclareMathOperator{\cm}{cm}  \DeclareMathOperator{\g}{g} \DeclareMathOperator{\erg}{erg} \DeclareMathOperator{\km}{km}  \DeclareMathOperator{\s}{s}    \DeclareMathOperator{\few}{few} 
       \newcommand{\cL}{{\cal L}} \newcommand{\cM}{{\cal M}}  \newcommand{\cO}{{\cal O}} \newcommand{\cP}{{ \cal P}}   \newcommand{\cS}{{\cal S}}
\newcommand{\ep}{\epsilon}
\newcommand{\epm}{\epsilon_{\rm m}}
  \newcommand{\eg}{{\it e.g.}}
   \def\oL{\overline} 
\newcommand{\pL}{\left(} \newcommand{\pR}{\right)} \newcommand{\bL}{\left[} \newcommand{\bR}{\right]} \newcommand{\cbL}{\left\{} \newcommand{\cbR}{\right\}} \newcommand{\mL}{\left|} \newcommand{\mR}{\right|}
\newcommand{\beq}{\begin{equation}} \newcommand{\eeq}{\end{equation}}
\newcommand{\bea}{\begin{eqnarray}} \newcommand{\eea}{\end{eqnarray}}
\newcommand{\alg}[1]{\begin{align} \begin{split} #1 \end{split}  \end{align}}
\newcommand{\vev}[1]{\langle {#1} \rangle}
\newcommand{\tenx}[1]{\times 10^{#1}}
\newcommand{\Eq}[1]{Eq.~(\ref{#1})} \newcommand{\Eqs}[2]{Eqs.~(\ref{#1}) and (\ref{#2})} 
\newcommand{\Sec}[1]{Sec.~\ref{#1}}  
\newcommand{\Fig}[1]{Fig.~\ref{#1}} \newcommand{\Figs}[2]{Figs.~\ref{#1} and \ref{#2}}

\newcommand{\App}[1]{App.~\ref{#1}} 
 \DeclareMathOperator{\tr}{Tr}  \DeclareMathOperator{\re}{Re} \DeclareMathOperator{\im}{Im}

\begin{document}
\title{Supernova 1987A Constraints on Sub-GeV Dark Sectors, Millicharged Particles, the QCD Axion, and an Axion-like Particle}
\author{Jae Hyeok Chang${}^a$, Rouven Essig${}^a$, and Samuel D.~McDermott${}^b$}
\affiliation{${}^a$ C.~N.~Yang Institute for Theoretical Physics, Stony Brook, NY, USA}
\affiliation{${}^b$ Fermi National Accelerator Laboratory, Center for Particle Astrophysics, Batavia, IL, USA}
\date{\today}

\abstract{ 
We consider the constraints from Supernova 1987A on particles with small couplings to the Standard Model. We discuss a model with a fermion coupled to a dark photon, with various mass relations in the dark sector; millicharged particles; dark-sector fermions with inelastic transitions; the hadronic QCD axion; and an axion-like particle that couples to Standard Model fermions with couplings proportional to their mass.  In the fermion cases, we develop a new diagnostic for assessing when such a particle is trapped at large mixing angles. 
Our bounds for a fermion coupled to a dark photon constrain small couplings and masses $\lesssim 200\mev$, and 
do not decouple for low fermion masses.  
They exclude parameter space that is otherwise unconstrained by existing accelerator-based and direct-detection searches.  
In addition, our bounds are complementary to proposed laboratory searches for sub-GeV dark matter, and do not constrain 
several benchmark-model targets in parameter space for which the dark matter obtains the correct relic abundance from interactions 
with the Standard Model.  
For a millicharged particle, we exclude charges between $10^{-9}-\few\tenx{-6}$ in units of the electron charge, also for masses $\lesssim 200\mev$; this excludes parameter space to higher 
millicharges and masses than previous bounds. 
For the QCD axion and an axion-like particle, we apply several updated nuclear physics calculations and include the energy dependence of the optical depth to accurately account for energy loss at large couplings. These corrections allow us to rule out a hadronic axion of mass between $0.1$ and a few hundred eV, or equivalently to put a bound on the scale of Peccei-Quinn symmetry breaking between a $\few\tenx4$ and $10^8$ GeV, closing the hadronic axion window.  
For an axion-like particle, our bounds disfavor decay constants between a $\few\tenx 5$ GeV up to a $\few\tenx 8$ GeV, for a mass $\lesssim 200\mev$. In all cases, our bounds differ from previous work by more than an order of magnitude across the entire parameter space.  We also provide estimated systematic errors due to the uncertainties of the progenitor. 
}

\preprint{YITP-SB-18-01, FERMILAB-PUB-17-432-T}

\maketitle

\setcounter{page}{2}

\section{Introduction}

In 1987, a core-collapse supernova known as Supernova 1987A (SN1987A) was observed in the Large Magellanic Cloud.  
SN1987A provided a wealth of information on the supernova explosion itself 
and also sets unique constraints on the existence of new, low-mass particles that are weakly-coupled to the 
Standard Model (SM)~\cite{Raffelt:1987yt,Raffelt:1996wa}. 

The existence of new, weakly-coupled particles could provide novel channels to ``cool'' the proto-neutron star and change the 
neutrino emission from SN1987A.  
Constraints in this context are derived from the ``Raffelt criterion'': the luminosity carried by the new particles from the interior of the proto-neutron star environment to the outside of the neutrinosphere must be smaller than the luminosity carried by neutrinos~\cite{Raffelt:1996wa}. 
The observed cooling time of the supernova agrees within uncertainties with the SM prediction~\cite{Burrows:1986me, Burrows:1987zz}. 
However, if there were an additional efficient channel for energy flow that could compete with neutrinos, the cooling time of the supernova would have been shorter than observed. 

SN1987A provides a hot and dense stellar environment, so even very weakly-coupled particles 
could have been produced. 
Since the supernova core temperature, $T_c$, is about 30~MeV, particles with mass less than about a few hundred MeV can be constrained
when taking into account the Boltzmann tail.  
In this paper, we will derive constraints from SN1987A on several possible low-mass particles: 
various dark sectors consisting of dark matter (DM) and 
dark photons (including millicharged particles), the QCD axion, and axion-like particles with Yukawa couplings. 

A popular and prototypical model for sub-GeV DM is given by a dark sector consisting of a DM particle, $\chi$, interacting with a 
dark photon, $A'$~\cite{Boehm:2003hm,Pospelov:2007mp,Hooper:2008im,ArkaniHamed:2008qn,Pospelov:2008jd,Feng:2009mn,Essig:2011nj}.  
Kinetic mixing between the dark photon and the SM photon leads to an interaction between the DM and electrically 
charged particles of the SM.  Various hierarchies between the DM mass, $m_\chi$, and the dark photon mass, $m'$, 
lead to a diverse range of phenomenology.  
In several cases, sharp benchmark targets can be identified in parameter space for which the DM interacting with a dark photon 
can obtain the correct 
DM relic density~\cite{Borodatchenkova:2005ct,Essig:2011nj,Chu:2011be,Izaguirre:2015yja,Essig:2015cda} 
(for a review, see~\cite{Alexander:2016aln,Battaglieri:2017aum}).  
In this paper, we will consider ``heavy'' DM with $m'<2m_\chi$, ``light'' DM with $m'>2m_\chi$, and a dark-sector with 
a millicharged particle (the latter bounds apply equally well to DM that interacts with a 
massive, but ultralight, $A'$).  In addition, we will consider ``light'', inelastic DM, 
in which the dark-sector consists of two ``DM'' particles, $\chi_1$ and $\chi_2$, which have a small mass splitting, and for which 
the interaction with the dark photon is off-diagonal, i.e.~$\sim A'\bar\chi_1\chi_2$.  

In deriving the SN1987A constraints on these particles, 
we include the thermal effects on $A'$-photon mixing~\cite{An:2013yfc,Redondo:2013lna}, 
which are very important for the SN1987A constraints on a dark sector consisting of only dark 
photons~\cite{Chang:2016ntp,Hardy:2016kme} 
(for previous bounds see~\cite{Mahoney:2017jqk,Bjorken:2009mm,Dent:2012mx,Kazanas:2014mca,Rrapaj:2015wgs}).  
As we will see, the presence of the DM particles changes the constraints in a significant way from the $A'$-only case, 
even when the $A'$ 
is kinematically forbidden to decay to the DM directly.  
We also use a novel criterion to calculate the couplings for which the DM is ``trapped'' inside the proto-neutron star (and thus does not 
contribute to the cooling):  we require them to take a random walk until in their velocity vector is turned by $90^\circ$ from their 
initial direction of motion. 
The SN1987A constraints on the light DM scenario had been considered previously 
in~\cite{Izaguirre:2013uxa,Essig:2013vha,Dreiner:2013mua,Izaguirre:2014bca}, but our analysis goes well beyond these references. 
Our analysis also significantly updates previous work on millicharged particles~\cite{Davidson:2000hf}. 

In addition to the dark-sector models mentioned above, we also revisit the constraints on another popular and important 
particle, the QCD axion~\cite{Preskill:1982cy, Abbott:1982af, Dine:1982ah}. 
Previous bounds have been extracted with a range of simplifying assumptions, which we attempt to rectify by including additional estimates of 
known nuclear physics as well as particle physics effects.  
We find significant differences with the constraints in the literature~\cite{Patrignani:2016xqp}.  
Finally, we also revisit constraints on axion-like particles with couplings proportional to the SM Yukawa couplings, 
updating bounds from~\cite{Essig:2010gu}.  

The remainder of the paper is organized as follows. 
In \Sec{sec:DM} we discuss the production of DM and dark photons and describe in some detail how we calculate the luminosity of these dark-sector particles.  We discuss the case of ``heavy'' and ``light'' DM, defined by whether the dark photon is lighter or heavier than 
twice the DM mass, respectively.  We also discuss variants of the basic DM-coupled-to-a-dark-photon model. This includes inelastic DM,  
in which case additional mass terms in the Lagrangian allow the fermion states to have different masses and off-diagonal couplings to the dark 
photon; and the case where the dark photon is essentially massless, so that the DM appears to have a ``millicharge.''
\Sec{sec:DM-results} describes the results for these various models.  
In \Sec{axionsection} we change gears entirely and address the QCD axion, while \Sec{alpsection} discusses axion-like particles.  
In \Sec{conclusionsection} we conclude.  
We leave many details of our calculations to various Appendices.

\section{Dark Matter Coupled to a Dark Photon: Model and Analysis}
\label{sec:DM}

\subsection{Model Description and Preliminary Comments}\label{subsec:models}

We consider a variation of the model examined in~\cite{Chang:2016ntp}, in which the only new light particle was a dark photon of a new $U(1)'$ gauge group that kinetically mixes with the SM hypercharge gauge boson.  
This dark photon is a massive vector boson with a small coupling to electrically charged particles.  
Here, we assume that the dark sector also includes a Dirac fermion, $\chi$, charged under the $U(1)'$, that is light enough to be produced in nucleon-nucleon collisions in the proto-neutron star. Thus, pair production of $\chi \bar \chi$ states provides an additional channel through which energy can flow. The low-energy Lagrangian describing such a dark sector is
\beq
\cL_{\rm dark} = - \frac14 F'_{\mu \nu} F'^{\mu \nu} - \frac\ep2 F'_{\mu \nu} F^{\mu \nu}  - \frac12 m'^2 A'_\mu A'^\mu + \bar \chi \pL i \gamma^\mu \partial_\mu + g_D \gamma^\mu A'_\mu -m_\chi  \pR \chi\,,
\eeq
where $\epsilon$ is the kinetic-mixing parameter, $g_D$ is the dark gauge coupling, $m'$ ($m_\chi$) is the dark photon ($\chi$) mass, $F_{\mu\nu}$ the usual electromagnetic field-strength tensor, and $F'_{\mu\nu}$ the field-strength tensor of the $U(1)'$ gauge boson.  
We define $\alpha_D\equiv g_D^2/4\pi$ as the ``dark fine-structure constant.'' 
Conservation of $\chi$-fermion number guarantees that these particles are stable even below the scale of $U(1)'$ symmetry breaking, which is why they are a DM candidate. For this reason, we refer to $\chi$ as ``DM'' in this paper, although we will not address its early-universe production nor its cosmological effects. Moreover, as long as $\chi$ is stable on the time it takes to escape the proto-neutron star, the SN1987A constraints derived below are applicable even if $\chi$ is only a fraction of the DM.  Of particular interest for understanding their behavior in SN1987A, conservation of $\chi$-number means that the dark fermions can scatter off SM particles as they make their way out of the star. 
This affects the energy spectrum of the $\chi$ particles and the $A'$ compared to the scenario with a solitary $A'$, and qualitatively changes the notion of a trapping limit at large mixing angle. 
Furthermore, $\chi \bar\chi$-pairs may be produced through off-shell dark photons, which can avoid the suppression from thermal effects on $A'$-photon mixing described in~\cite{Chang:2016ntp}. As a result, the lower bounds have a different low-mass limiting behavior and become stronger compared to the $A'$-only case.

Our constraints are derived explicitly for a dark sector with a Dirac fermion coupled to the dark photon. 
However, the constraints should be very similar for a dark sector consisting of a complex scalar coupled to the dark photon.  
The production and (relativistic) scattering cross sections are slightly modified, but the particle number is still conserved, so the kinematics of scattering are similar.
We expect differences due to degree of freedom counting and details of the cross sections to give $\mathcal{O}(1)$ corrections to 
the fermionic bounds. 
These differences are below the systematic uncertainties due to imperfect knowledge of the supernova 
temperature and density profiles, discussed in more detail below. Thus, our bounds can serve as rough guidelines on the parameter space of a dark photon coupled to a dark charged scalar. 

In what follows, we ignore the presence of a dark Higgs boson, which can affect the phenomenology if the $U(1)'$ is broken through a dark 
Higgs mechanism and the Higgs boson remains very light.  The dark Higgs mass is determined by its self-coupling parameter, which we will henceforth take to be large enough that the dark Higgs is heavier than the dark photon and DM and kinematically inaccessible during the supernova. 
However, while there are additional dark-sector production and decay modes to consider, we claim that these should not lead to a 
significant change in the supernova cooling rate compared to the rates we derive here. Processes involving dark Higgs production do not suffer 
suppression from the well-known plasma effects in on-shell $A'$-photon mixing, but such suppression is also absent for DM-pair production. Thus,
the $A'$-only case explored in~\cite{Chang:2016ntp} is in some sense unique, and the results in this work
should be qualitatively similar even with a light dark Higgs boson.

As we discuss in more detail below, there are several important processes by which DM particles are created in the proto-neutron star environment. The dominant processes at low DM mass are bremsstrahlung of a DM pair during nucleon collisions through an on-shell or off-shell $A'$ and SM photon decay to a DM pair in the plasma if $m_\chi \lesssim \omega_{p,0}$, where $\omega_{p,0} \equiv \omega_p(r=0) \sim 15 \mev$ is the plasma frequency at the center of the supernova (see \App{aprime-prod}).
After being produced, DM particles scatter against nucleons and electrons on their way out of the proto-neutron star environment.
Similar to the $A'$-only case, where the rate of $A'$ production as well as decay or absorption are proportional to the same parameter ($\ep^2$), 
the rate at which the DM particles are produced and the rate at which they scatter are proportional to the same parameter combination ($\alpha_D\ep^2$).
At increasingly large mixing angle, the DM particles will scatter multiple times during egress, potentially becoming trapped and even returning to chemical equilibrium inside the proto-neutron star. 
We find that there is some parameter space where $\alpha_D \ep^2$ is large enough that a sufficient number of dark fermions are produced to alter the evolution of the supernova explosion but small enough that the dark-sector particles scatter infrequently on their way out of the star. Thus, for a large range of DM masses, there is both a lower and upper bound on the coupling to the SM.

We emphasize here that if the elastic scattering of DM particles is extremely frequent, they may be unable to exit the supernova. If $\chi$ and $\bar \chi$ particles proliferate throughout the star, DM annihilation or pairwise inverse bremsstrahlung will equilibrate everywhere, attaining a thermal abundance on a timescale set by their production rate~\cite{Weldon:1983jn}.\footnote{We thank N.~Toro for useful discussions.} 
Equipartition of degrees of freedom in the thermal bath then implies that the overall temperature as well as the transport properties of the proto-neutron star will change. Although introducing this many new degrees of freedom may have unacceptable consequences for the behavior of the supernova explosion, such effects are hard to resolve analytically and potentially beyond the present day understanding of the proto-neutron star interior, and thus are beyond the scope of this work. Here we will calculate the mixing angle that gives a decoupling radius close to the neutrinosphere using conservative analytic requirements, and we assume a benign thermal population at higher mixing angles. We will show these as plausible upper bounds for the kinetic mixing parameter above which the effects of the dark sector are best investigated with other, laboratory-based techniques, as discussed in e.g.~\cite{Battaglieri:2017aum}. 
It would certainly be interesting to use simulations to investigate more comprehensively the effects of a
thermally-equilibrated dark sector population on supernova explosions.

\subsubsection{Model Variation: Inelastic Dark Matter}\label{subsubsec:iDM-model}

The model above describes a Dirac fermion coupled to a dark photon.  An economical, UV-complete model of dark-sector masses is provided by the introduction of a doubly-charged dark Higgs field $h_D$ that experiences spontaneous symmetry breaking with a nonzero vacuum expectation value $\vev{h_D}$. If the $U(1)'$-charged fermions have Yukawa couplings to the dark Higgs as well as a $U(1)'$-invariant Dirac mass, the mass eigenstates can undergo a small splitting after symmetry breaking. The Lagrangian for this scenario is
\beq
\mathcal{L} \supset \frac{i}2 \lambda^\dagger \bar{\sigma}^\mu D_\mu \lambda + \frac{i}2 \xi^\dagger \bar{\sigma}^\mu D_\mu \xi + m_\chi \lambda \xi + m_\lambda \lambda \lambda + m_\xi \xi \xi + h. c.,
\eeq
where $\chi=(\lambda ~~ \xi^\dagger)$; $\lambda, \xi$ are two-component Weyl fermions; and $m_\xi, m_\lambda \propto \vev{h_D}$ arise after symmetry breaking. A field redefinition allows us to work in terms of mass eigenstate fields $\chi_1, \chi_2$. These fermions have different masses and in principle can couple to the dark photon inelastically as well as elastically:
\alg{
\mathcal{L} \supset& \frac{i}2 \chi_1^\dagger \bar{\sigma}^\mu D_\mu \chi_1 + \frac{i}2 \chi_2^\dagger \bar{\sigma}^\mu D_\mu \chi_2 + \frac{m_1}2 \chi_1 \chi_1 + \frac{m_2}2 \chi_2 \chi_2 \\
   & + g_D \bL \frac{2 i m_\chi}{M} \chi_1^\dagger \bar{\sigma}^\mu \chi_2 + \frac{m_\xi-m_\lambda}{2M} \pL \chi_1^\dagger \bar{\sigma}^\mu \chi_1 + \chi_2^\dagger \bar{\sigma}^\mu \chi_2 \pR \bR A'_\mu + {\rm (h.\,c.)},
}
with
\beq
M^2 = 4m_\chi^2+(m_\xi-m_\lambda)^2, \qquad m_{1,2}=\frac{1}{2} \bL M \mp (m_\xi + m_\lambda) \bR.
\eeq
We define the mass splitting $\Delta \equiv m_2-m_1=m_\xi + m_\lambda$.  The lighter of the two Majorana fermions, $\chi_1$, could be a DM 
candidate.  We refer to this model as ``inelastic DM''~\cite{TuckerSmith:2001hy}, 
since the scattering of $\chi_1$ off SM particles could be dominated by a transition 
from $\chi_1$ to the heavier state $\chi_2$.  This model was studied in~\cite{Izaguirre:2015zva,Izaguirre:2017bqb} in 
the sub-GeV mass range of interest in this paper. 
If $m_\xi=m_\lambda$ exactly, 
the elastic coupling vanishes and only inelastic scattering is possible at tree-level.  
We will calculate below the SN1987A constraints for this simple variant of the inelastic DM model.  
We note that even in this case $\chi_1$ could scatter elastically at one-loop, although this is highly suppressed, 
as we will discuss further in \Sec{subsec:results-iDM}.  

\subsubsection{Model Variation: Millicharged Particles}\label{subsubsec:milli-model}

A millicharged particle is a dark-sector particle with a small electric charge.  One well-motivated way to attain such small charge is by introducing a massless dark photon, which can be removed by a field redefinition of the SM photon.  Under such a field redefinition, dark-sector particles coupled to the $A'$ acquire a small coupling to the SM photon, $\epsilon g_D$, where $\epsilon$ is again the kinetic-mixing parameter between SM photon and dark photon. We define $Q \equiv \epsilon g_D/e$ so that dark-sector particles charged under the dark photon have an electric charge $Q e$.  

We will derive below the constraints from SN1987A on millicharged particles, which update bounds presented in~\cite{Davidson:2000hf}.  
These bounds are equally applicable for DM particles that couple to an ``ultralight'', massive dark-photon mediator, with a mass below the typical momentum transfer in DM production and scattering processes, i.e.~so long as the mediator mass can be neglected in 
the calculations~\cite{Essig:2015cda}. 

\subsection{Dark-Sector Particle Production in the Proto-Neutron Star}
\label{darkmatterinsn}

\begin{figure}[t]
\begin{center}
\includegraphics[width=.56\textwidth]{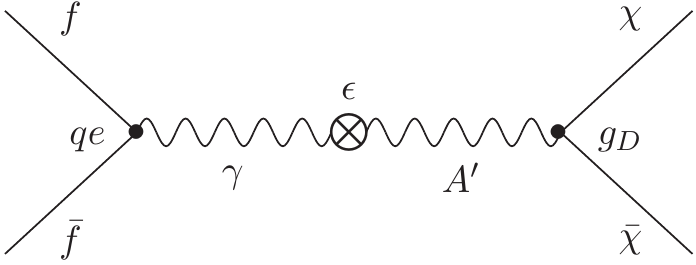}
\caption{The Feynman diagram for the interaction between $\chi$ and Standard Model particles $f$, which have a charge $q e$, where $e$ is the 
electron's charge.}
\label{feyn-diags-int}
\end{center}
\end{figure}

\begin{figure}[t]
\begin{center}
\includegraphics[width=0.95\textwidth]{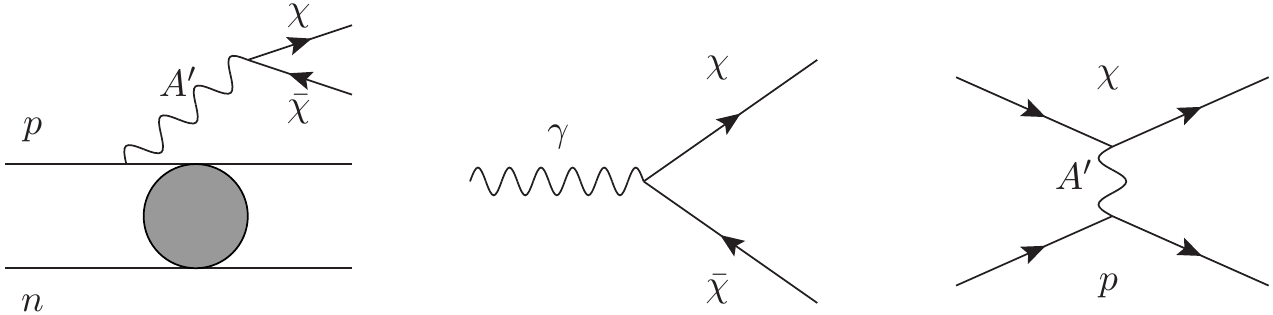}~
\caption{Processes relevant for the dark matter in the interior of the proto-neutron star. In the left panel, the $A'$ can be on- or off-shell. 
The coupling of the $A'$ to the Standard Model particles is through kinetic mixing with the photon as shown in 
\Fig{feyn-diags-int}. }
\label{feyn-diags}
\end{center}
\end{figure}

Dark photons and particles charged under $U(1)'$ can be produced in the proto-neutron star through the kinetic mixing between the SM photon and the dark photon, see \Fig{feyn-diags-int}.
The total luminosity in dark-sector particles is $L_{\rm dark} = L_\chi + L_{A'}$ (note that $L_\chi$ denotes the total luminosity in $\chi$ and $\bar\chi$), and the criterion $L_{\rm dark} = L_\nu$ determines the boundary of constraints, where $L_\nu = 3\tenx{52}\erg/\s$ is the neutrino luminosity at one second~\cite{Raffelt:1996wa}. The dominant production mechanisms for DM particles are through bremsstrahlung (via an off- or on-shell $A'$) during neutron-proton collisions and through SM photon decay in the plasma, shown in the left and middle panels, 
respectively, of \Fig{feyn-diags}.  
For on-shell $A'$ production, bremsstrahlung dominates, while for $\chi$ production both bremsstrahlung and SM photon decay are important, 
with the latter dominating by a factor of a few.

Before discussing the calculation of $L_{\rm dark}$, it is worth making some general 
comments on the DM production in the supernova.  
We then describe the calculation of $L_{A'}$ 
and $L_\chi$ in \Sec{subsec:LA'} and \Sec{subsec:Lchi}, respectively, leaving detailed formulae to the Appendices.  

Due to plasma effects in the proto-neutron star interior, we find it convenient to calculate scattering amplitudes in the gauge boson interaction basis, as in~\cite{An:2013yfc}. With this choice, all Feynman diagrams describing the interaction of DM with electrically charged particles such as the proton implicitly contain the diagram of \Fig{feyn-diags-int}, for which the amplitude is
\alg{ \label{med-amplitude}
\mathcal{M}&=-e J^\mu_{em} \left< A_\mu A_\nu \right> \epsilon K^2 g^{\nu\rho} \left< A'_\rho A'_\sigma \right> g_D J^\sigma_{\chi} 
\\ &= \epsilon e g_D \left( \frac{K^2}{K^2-m'^2+i m' \Gamma' + \Pi_D} \right) J^\mu_{em} \left(\frac{\cP_{T\mu\nu}}{K^2-\Pi_T}+\frac{\cP_{L\mu\nu}}{K^2-\Pi_L} \right) J^\nu_{\chi} \,,
}
where $K_\mu=(\omega,\vec{k})$ is the momentum four-vector of the intermediate state (carried by both the SM photon and the dark photon), $\Pi_D$ is the self-energy of the dark photon in a plasma of DM particles, $\cP_{T\mu\nu}$ and $\cP_{L\mu\nu}$ are the transverse and longitudinal projection operators of the SM polarization states, respectively, and $\Pi_T$ and $\Pi_L$ are polarization tensors of the SM photon from thermal effects. The dark photon absorptive width $\Gamma'$ is dominated by its decay width to DM when this decay is on shell,
\beq
\Gamma' \simeq \frac{\alpha_D m'}3 \sqrt{1-\frac{4m_\chi^2}{m'^2}}\pL 1+\frac{2m_\chi^2}{m'^2} \pR \Theta \pL m'-2m_\chi \pR + \cO(\ep^2) \equiv \Gamma_\chi +\cO(\ep^2) \, .
\eeq
In principle, the presence of the dark photon self-energy $\Pi_D$ suggests that we should include separate longitudinal and transverse projection operators for the dark photon like we do for the SM photon. However, $\Pi_D$ is negligible on the lower boundary of the excluded parameter space where we calculate dark-sector production rates, since the dark-sector particles free stream. The effect of $\Pi_D$ is only important near the upper boundary where the dark-sector number densities can be very high.  However, as we discuss in more detail below, in this part of parameter space it is safe to assume that number densities are simply given by a thermal distribution inside some radius, so their exact production rate will not be important. Thus, we will not use $\Pi_D$ in any explicit calculation.

The dark sector luminosity admits two kinds of resonances, as can be seen in \Eq{med-amplitude}: the on-shell peak from the dark photon propagator, attained for $K^2 = m'^2$, and the ``thermal peak,'' at $K^2 = \Pi_{L,T}$ from the SM photon propagator. The on-shell peak dominates if $m' \gg \omega_{p,0}$, the thermal peak dominates if $m' \ll \omega_p$, and both peaks can be attained for $m' \sim \omega_p$. Thus for $m' \gtrsim \omega_p$, off-shell DM production is suppressed, and the lower bounds are same as the dark photon only case. However, off-shell DM production dominates for $m' \ll \omega_{p,0}$, so the low-$\ep$ bounds at small $m'$ are stronger than the bound from the $A'$-only case, which decouples like $m'^2$ due to suppression by thermal effects~\cite{Chang:2016ntp, An:2013yfc}. With the inclusion of dark sector fermions, the production rate for dark-sector particles becomes independent of their mass, so the lower bound is flat.

\subsection{Dark Photon Luminosity ($L_{A'}$) for Small Couplings}\label{subsec:LA'}

We now discuss the calculation of the luminosity $L_{A'}$ for small couplings $\ep$ and $\alpha_D$. 
This is similar to the pure $A'$ case discussed in~\cite{Chang:2016ntp} 
only if the DM particles free stream out of the proto-neutron star.  
If, however, the couplings are large, then the dark matter abundance is also large and the $A'$ will experience a large optical depth. 
We will discuss our treatment of the bounds at large coupling values in \Sec{subsec:trapping};  \Eq{blackbody-boson} gives the $A'$ luminosity for 
large couplings. 
In this section, we will ignore dark-sector interactions. 

Bremsstrahlung production of the $A'$ dominates over purely electromagnetic-like processes such as semi-Compton scattering, because the
QCD coupling $\alpha_s \sim \cO(4\pi) \gg \alpha_{\rm EM} \simeq 1/137$ and because nucleons are highly abundant but not Pauli blocked like electrons. 
SM fermion annihilation contributes negligibly because the chemical potential of all such particles is very high and their antiparticles are very scarce.
Mixed nucleon scattering dominates over proton-proton scattering because the former emits dipole radiation while the latter only emits like a quadrupole~\cite{Nyman:1968jro, Rrapaj:2015wgs}.
The ``direct luminosity'' in $A'$ particles is 
\alg{ \label{L-A'}
L_{A'} &= \int_0^{R_\nu} dV \int \frac{d^3 \vec k}{2 \omega (2\pi)^3} e^{-\tau(\omega, r) } \omega \Gamma_{\rm br}(\omega,r)\,,
\\ \tau(\omega, r)&= \int_r^{R_{\rm far}} dr' \bL \Gamma_{\rm ibr}(\omega,r') + \Gamma_e(\omega,r') + \Gamma_\chi(\omega,r') + \Gamma_{\rm dC}(\omega,r') \bR ,
}
where $R_\nu$ is the neutrinosphere radius, $\tau$ is the optical depth, $\Gamma_{\rm br}$ is the dark photon production rate via bremsstrahlung, 
$\Gamma_{\rm ibr}$ is the inverse bremsstrahlung rate, $\Gamma_e$ ($\Gamma_\chi$) is the $A'$ decay width to electromagnetic (DM) particles, and $\Gamma_{\rm dC}(\omega,r')$ is the rate for ``dark Compton'' scattering (e.g.~$A' \chi \to A' \chi$) 
that only contributes when the $\chi$ are trapped (see \Sec{subsec:trapping}). The far radius $R_{\rm far}$ is taken to be the neutrino gain radius $R_g \simeq 100\km$~\cite{Chang:2016ntp, Bethe:1992fq}.
\App{aprime-prod} contains the definitions of all these rates.

In principle, DM annihilation as well as semi-Compton scattering involving a single SM photon and a single dark photon can also 
contribute to the power, but these are negligible unless the DM is trapped, which we explicitly ignore for the time being. 
The widths $\Gamma_{\rm ibr}$ and $\Gamma_e$ are suppressed by $\ep^2$, and $\Gamma_\chi$ is nonzero only if $m'>2m_\chi$.
If $m'>2m_\chi$ dark photons decay to DM particles on very short distances compared to the size of the proto-neutron star (unless $\alpha_D$ is 
very small), which sends $e^{-\tau(\omega, r)} \to 0$ such that $L_{A'}\simeq 0$ and $L_{\rm dark} \simeq L_\chi$.

\subsection{Dark Matter Luminosity ($L_\chi$) for Small Couplings}\label{subsec:Lchi}

There are two main contributions to the DM production rate: (i) bremsstrahlung of DM pairs in proton-neutron collisions, which we call $L_\chi^b$, and 
(ii) SM photon decays in the thermal plasma, which we call $L_\chi^d$; see left and middle panel of \Fig{feyn-diags}.  
Assuming that the DM particles do not scatter on their way out of the star (valid for small values of $\alpha_D\ep^2$), it is straightforward to 
calculate the resulting dark-fermion luminosity, $L_\chi = L_\chi^b+L_\chi^d$.  
We will give the corresponding expressions 
in \Sec{subsubsec:brems} and \Sec{subsubsec:photon-decay}.  

For large values of $\alpha_D$ and $\ep$, 
the DM particles may scatter and thermalize with SM material, rendering their escape energy different from their 
energy at production.  Since DM number is conserved, the dark fermion luminosity then does not have a description analogous to \Eq{L-A'}.
Moreover, as mentioned in \Sec{subsec:LA'}, the dark-photon luminosity $L_{A'}$ needs to be modified from \Eq{L-A'} in the presence of DM 
particles and large values of $\alpha_D$ and $\ep$, 
since in this case the $A'$ can scatter off DM particles, which conserves $A'$ number.  
For large couplings, the calculations of $L_\chi$ and $L_{A'}$ then require defining a ``trapping criterion''.  
Various choices for such a definition are in principle possible; we describe our criterion in \Sec{subsec:trapping}.

\subsubsection{Bremsstrahlung of Dark Matter Pairs}\label{subsubsec:brems}

DM pairs can be produced by bremsstrahlung in proton-neutron collisions.  If the DM particles do not scatter on their way out of the star, the  differential luminosity per unit volume is
\alg{ \label{diffP-DM}
\frac{d L^b_\chi}{dV}&= \int d\Pi_{p_1} f_1 d\Pi_{p_2} f_2 d\Pi_{p_3} (1-f_3) d\Pi_{p_4} (1-f_4) d\Pi_{p_\chi} (1-f_{\chi}) d\Pi_{p_{\bar\chi}} (1-f_{\bar\chi})\times \\
& \qquad\qquad\qquad\qquad\qquad \times (2\pi)^4\delta^4(p_1+p_2-p_3-p_4-p_\chi-p_{\bar\chi}) (E_\chi+E_{\bar{\chi}}) |\mathcal{M}|^2,
}
where $d\Pi_{p_i} =\frac{d^3 \vec p_i}{2E_i(2\pi)^3}$ is the Lorentz-invariant phase space of the particle with four-momentum $P_\mu =(E_i, \vec p_i)$, the nucleon phase space densities are $f_i$, and the incoming (outgoing) nucleons have momenta $p_1, p_2$ ($p_3, p_4$). As in~\cite{Chang:2016ntp} we employ the soft radiation approximation to calculate the matrix element, which is valid for the mass range in which $E_\chi + E_{\bar\chi} \ll |\vec p_N|^2/2m_N$~\cite{Rrapaj:2015wgs}. Details of the calculation of \Eq{diffP-DM} are in \App{chi-prod}. The luminosity due to emission of DM in this limit is the volume integral of \Eq{diffP-DM}, $L^b_\chi = \int_0^{R_\nu}dV \frac{dL^b_\chi}{dV} \,,$ assuming that dark matter particles free stream out of the star.

\subsubsection{Standard-Model Photon Decay in the Thermal Plasma}\label{subsubsec:photon-decay}

In vacuum, dark-sector particles charged under $U(1)'$ do not couple to on-shell SM photons because the mixing term is $\epsilon K^2$ and $K^2=0$ for on-shell photons. Since the dispersion relation for photons is altered in a thermal plasma, however, this coupling does arise. The SM photon dispersion relation picks up a real part $\re(\Pi_{L,T}) \leq \sqrt{3/2} \ \omega_p$ 
(where $\omega_p$ is a function of the distance from the center of the proto-neutron star, $r$) 
and a nonzero imaginary part. DM with mass less than $\sqrt{3/2}\ \omega_p/2$ can therefore be produced from the decay of SM photons, shown in \Fig{feyn-diags}, much like the plasmon process that leads to neutrino production~\cite{Braaten:1993jw}. 

We can write the differential luminosity of the DM from SM photon decay as 
\beq \label{SM-photon-decay-lum}
\frac{dL^d_\chi}{dV}= \int \frac{d^3\vec k}{(2\pi)^3} \pL \frac{2 \omega_T \Gamma^d_T}{e^{\omega_T/T}-1}+\frac{\omega_L \Gamma^d_L}{e^{\omega_L/T}-1} \pR,
\eeq
where $K_{T,L}^\mu = (\omega_{T,L}, \vec k)$ is the photon four-momentum and $\Gamma^d_{T,L}$ is its decay rate to a DM pair
\alg{ \label{SM-photon-decay-wid}
\Gamma^d_{T,L}&=\frac{1}{2\omega_{T,L}} \int \frac{d^3 \vec p_\chi}{(2\pi)^3 2E_\chi} \int \frac{d^3 \vec p_{\bar{\chi}}}{(2\pi)^3 2E_{\bar{\chi}}} (2\pi)^4 \delta^4(K^\mu-P_\chi^\mu-P_{\bar\chi}^\mu) |\mathcal{M}^d_{T,L}|^2\,,
\\ |\mathcal{M}^d_T|^2&=8\pi\alpha_D \epsilon^2 \frac{K^4(K^2-2p_\chi^2 \sin^2\theta_\chi)}{(K^2-m'^2)^2+(m' \Gamma_\chi)^2}, ~~ |\mathcal{M}^d_L|^2=8\pi\alpha_D \epsilon^2 \frac{K^4 \bL K^2-4(P_\chi^\mu \epsilon_{L \mu})^2 \bR}{(K^2-m'^2)^2+(m' \Gamma_\chi)^2}\,.
}
Here, $P^\mu_\chi = (E_\chi, p_\chi)$ is the four-momentum of the outgoing DM particle, $\epsilon_L^\mu=\pL \frac{k}{\sqrt{K^2}},\frac{\omega_L}{\sqrt{K^2}} \frac{\vec{k}}{k} \pR$ is the longitudinal polarization vector, and $\theta_\chi$ is the angle between the incoming photon and the outgoing DM. The amplitudes in \Eq{SM-photon-decay-wid} are not Lorentz invariant because the plasma frame breaks Lorentz invariance, and the transverse and longitudinal modes have different dispersion relations. In \Eq{SM-photon-decay-lum}, the momentum integral of the longitudinal mode is cut off at $k_{\textup{max}}$, the largest three-momentum that the longitudinal photon can have, while transverse photons can have any value of $k$~\cite{Braaten:1993jw}.

The luminosity of DM from Standard Model photon decay is a factor of a few higher than from bremsstrahlung for the mass range in which photon decay is possible; it is also independent of $m_\chi$ for $m_\chi \ll \omega_p$.
The luminosity due to emission of DM in this limit is the volume integral of \Eq{SM-photon-decay-lum}, $L^d_\chi = \int_0^{R_\nu}dV \frac{dL^d_\chi}{dV} \,,$ assuming that dark matter particles free stream out of the star.

\subsection{Dark-Sector Luminosity and Trapping Criterion for Large Couplings}
\label{subsec:trapping}

If the DM scatters many times on its way out of the star, it can equilibrate with the SM particles.
If the DM is in equilibrium, dark photons will be trapped as well. It is therefore important to define a decoupling radius $R_d$ for which DM is in equilibrium inside and free-streaming outside.\footnote{In fact, there are two ``decoupling radii,'' one each for chemical and kinetic equilibrium. The chemical decoupling radius $r_{\rm cd}$ determines the {\it number density} of DM particles emitted from the supernova, and the kinetic decoupling radius $r_{\rm kd}$ determines the {\it energy} of each DM particle being emitted. The chemical decoupling radius should be smaller than the kinetic decoupling radius because kinetic equilibrium is necessary for chemical equilibrium in this model. However, these radii are difficult to find exactly without simulations. As a conservative assumption, appropriate considering all the other uncertainties of the problem, we solve only for the radius of kinetic equilibrium. This is conservative because it provides a lower limit on the number density of DM particles, since $r_{\rm cd} \leq r_{\rm kd}$ in reality.} For a given $m_\chi$ and $m_{A'}$, this radius is obviously a function of $\ep$ and $\alpha_D$. We shall assume that dark sector energy emission from $R_d$ is free-streaming and thermal and the luminosity is an effective blackbody. We set the upper bound of $\epsilon$ to be where blackbody emission from $R_d(\alpha_D,\ep)$ equals $L_\nu$; we specify an algorithm for computing $R_d(\alpha_D,\ep)$ below. For larger values of $\alpha_D$ or $\epsilon$, the decoupling radius increases and the energy emission decreases because temperature falls sharply with radius. 
This is reminiscent of the calculations done for sterile neutrino emission from SN1987A~\cite{Kainulainen:1990bn}. 

The blackbody luminosity of a fermion from a radius $R_d$ is
\beq \label{blackbody-fermion}
\left. L_\chi(R_d) \mR_\text{therm.}=4 \pi R_d^2 \int dp_\chi \frac{g_\chi}{8 \pi^2}\frac{v_\chi E_\chi p_\chi^2}{e^{E_\chi/T(R_d)}+1} \underset{m_\chi \to0}\Longrightarrow \frac{7g_\chi \pi^3}{240}R_d^2 T(R_d)^4,
\eeq
where $E_\chi=\sqrt{p_\chi^2+m_\chi^2}, v_\chi=p_\chi/E_\chi$, $g_\chi=4$ counts degrees of freedom, and the final approximation assumes a massless fermion. The blackbody luminosity of a dark photon is
\beq \label{blackbody-boson}
\left. L_{A'}(R_d) \mR_\text{therm.}=4 \pi R_d^2 \int dp_{A'} \frac{g_{A'}}{8 \pi^2}\frac{v_{A'} E_{A'} p_{A'}^2}{e^{E_{A'}/T(R_d)}-1} \underset{m_{A'} \to0}\Longrightarrow \frac{g_{A'} \pi^3}{30}R_d^2 T(R_d)^4,
\eeq
and $\left. L_{\rm dark}(R_d) \mR_\text{therm.}=\left. L_{A'}(R_d) \mR_\text{therm.}+\left. L_\chi(R_d) \mR_\text{therm.} $. 
To derive a constraint on $\ep$ (given some choice for $\alpha_D$ and other model parameters), 
we (i) calculate the radius $R_d^*$ at which the dark sector blackbody luminosity equals the neutrino luminosity, $\left. L_{\rm dark}(R_d^*) \mR_\text{therm.}=L_\nu$, and then (ii) find the value of $\ep$ that gives thermal decoupling at this radius. 
Step (i) is computationally straightforward given that \Eqs{blackbody-fermion}{blackbody-boson} are simple to compute for a given mass.  
However, step (ii) is more involved, and we discuss our approach next. 

Since the ``decoupling radius'' is only an approximate concept and finding the zone of decoupling is impossible without simulations, we propose a simple criterion: the kinetic decoupling radius $R_d^*$ is defined to be where the {\it expected angular deflection} of a thermal DM particle starting at $R_d^*$ and ending at $R_f$ is $\pi/2$. At smaller radii $r<R_d^*$ most DM particles are redirected and find antiparticles to annihilate with, while at larger radii most DM particles escape and drain energy from the supernova explosion. To calculate the expected angular deflection, we first define the total number of scatters and the maximum angular deflection that a DM particle would experience if it was scattered in the same plane and in the same direction 
every time:
\beq \label{expected-deflection}
N (\alpha_D,\ep, R_d) =\int\limits_{R_d}^{R_f} \frac{dr \, \Gamma_s(\ep, \oL{E}(R_d), r) }{v} \,, ~~~ \theta_{\rm max}(\alpha_D,\ep, R_d) =\int\limits_{R_d}^{R_f} \frac{dr\, \Gamma_s(\ep, \oL{E}(R_d), r)  \Delta \theta }v\,.
\eeq
Here, $\Gamma_s$ is the event rate for $\chi+p \rightarrow \chi+p$ elastic scattering, $\Delta \theta $ is the average angular deflection per collision, and $\oL E(R_d)$ is the thermally averaged energy at the radius $R_d$. 
(We ignore the energy change of the DM after each collision, since it is small.  We also ignore the difference in total path length due to the angle change,  since this is a higher order effect.) Most particles will not move in the same angular direction upon each scattering, of course, but instead will take a random walk in solid angle. The expected displacement due to a random walk is the mean of the chi distribution in $d$ dimensions, which differs from the root-mean-square deviation $\theta_{\rm max}/\sqrt{N_{\rm steps}}$ by a factor $\sqrt{2/d}\times \Gamma\bL \pL d+1 \pR/2 \bR/\Gamma\bL d/2 \bR$. The expected angular deflection for a typical particle is thus
\alg{ \label{random-walk}
\left< |\theta(\alpha_D, \ep, R_d)|  \right> =\frac{\theta_{\rm max}(\alpha_D, \ep, R_d) }{2}& \sqrt{\frac{\pi}{N(\alpha_D, \ep, R_d) }} \implies
\\ \text{upper bound } &\ep_u  \text{ given by solving} ~~ \langle |\theta (\alpha_D, \ep_u, R_d^*)| \rangle = \frac\pi2.
}
Both $\theta_{\rm max}(\alpha_D, \ep, R_d)$ and $N(\alpha_D, \ep, R_d)$ scale like $\sim\!\alpha_D\ep^2$, so the expected angular deflection and the upper bound given in \Eq{random-walk} rise linearly in $\sqrt{\alpha_D}\ \ep$; as a result, if we choose a different critical angular deflection, for example $\left< |\theta(\alpha_D, \ep_u, R_d^*) | \right> =\pi$ instead of $\pi/2$, our bounds would be twice as restrictive. Details of the calculation of \Eq{expected-deflection} are given in \App{DMscat}.

We emphasize that our approach to calculating a constraint for large values of $\ep$ is quite different when DM 
is present compared to the $A'$-only case. In~\cite{Chang:2016ntp} we calculated the dark photon energy emission by weighting the differential power from all radii with the probability of escape, $e^{-\tau}$, regardless of the value of $\ep$. Since both the power and $\tau$ have nontrivial energy dependence, we found that the luminosity in dark photons is dominated by higher energies at higher mixing angle, though for large enough $\ep$ the Boltzmann suppression becomes important and the total luminosity decreases. This calculation is valid because dark photons {\it do not survive} scattering with SM particles. 
However, DM particles can elastically scatter many times and still escape, as can a dark photon that scatters off of DM. 
At large mixing angles, the dark sector energy distribution at escape may therefore be different from the energy distribution at production, unlike in \cite{Chang:2016ntp}. 
We also note that the DM elastic scattering cross section can be forward peaked if the mediator is light compared to the typical momentum transfer (a few MeV at the supernova core), so calculating the mixing angle at which $\int_{R_d}^{R_\nu} dr \Gamma_s \simeq1$ is misleading, as it is reasonable to expect that the DM can scatter at least once on its way out of the proto-neutron star without returning to chemical equilibrium.  
Our more involved calculation is necessary to obtain accurate limits.  

When calculating the upper boundary for large DM masses for the inelastic DM scenario, we will revert to a much 
simpler criterion than the one discussed above: we will require, very conservatively, that the DM scatters only once.  
We will further explain and justify this approach in \Sec{subsec:results-iDM}.

Finally, we note that for values of $\ep$ above our upper boundaries, DM inside of the proto-neutron star attains a thermal abundance out to radii larger than $R_d^*$. It is possible that some DM particles are able to escape and travel to the detectors that registered the SN1987A neutrinos. Upon arriving, they may be observed through elastic scattering with the water in the neutrino detectors. However, by assumption the thermal energy of DM particles at these masses is $\sim \cO(T(R_d)) \lesssim 3\mev$, which is below the threshold of the Kamiokande  detector~\cite{Hirata:1987hu}, so that only a few of the emitted particles far along the Boltzmann tail are detectable even in principle.  
In addition, unless the DM mass is very small, their arrival at the detector will be significantly delayed compared to the 
neutrino signal, and there will be a large spread in arrival times due to a large velocity dispersion.

\subsection{Supernova Temperature and Density Profiles}\label{subsec:profiles}

The calculations of dark-sector particle production and their luminosity require knowledge of the temperature and density profiles of the 
proto-neutron star.  There are large uncertainties in these profiles, and we thus use four different profiles in order to estimate the systematic 
uncertainties in our resulting constraints.  We choose the same profiles as in~\cite{Chang:2016ntp}, and will refer to these as the ``fiducial'' \cite{Raffelt:1996wa},
``Fischer, $11.8 M_\odot$'' \cite{Fischer:2016cyd},  ``Fischer, $18 M_\odot$'' \cite{Fischer:2016cyd},  and ``Nakazato, $13 M_\odot$'' \cite{Nakazato, Nakazato:2012qf}. 
The profiles from \cite{Fischer:2016cyd} use the AGILE-BOLTZTRAN code 
\cite{Hempel:2009mc,Liebendoerfer:2000fw,Liebendoerfer:2001gu,Liebendoerfer:2002xn,Liebendoerfer:2003es},
while the profile from \cite{Nakazato, Nakazato:2012qf} is based on a solution to a neutrino radiative hydrodynamical code before shock revival and a solution to the flux-limited diffusion equation after cooling has commenced.
See \cite{Chang:2016ntp} for further details and comparisons of these simulations. 
(See also~\cite{Kushnir:2014oca,Kushnir:2015mca,Blum:2016afe} for a qualitatively different explanation of the observed neutrino burst.)

\section{Dark Matter Coupled to a Dark Photon: Results}
\label{sec:DM-results}

The phenomenology of DM particles interacting with dark photons in the supernova depends on whether or not the dark photon can 
decay to DM, so we have to consider the two possible mass hierarchies between the $A'$ and $\chi$  separately. We thus study two scenarios: $\mathbf{(i)}$~``{\bf heavy dark matter}'', where we choose the specific mass relation $m_\chi=3m'$ for illustration, and $\mathbf{(ii)}$~``{\bf light dark matter}'', where we choose the specific mass relation $m'=3m_\chi$ for illustration.  In the former case, the dark photon can be stable against decays on the supernova timescales, since the decay is suppressed by $\epsilon^2$; in the latter case, all dark photons promptly decay to DM (we will assume that $\alpha_D$ is large enough to allow this). The constrained parameter space is very similar in both cases regardless of the mass hierarchy, since the luminosity is approximately a blackbody at large couplings, see \Eqs{blackbody-fermion}{blackbody-boson}. Even if the $A'$ is stable against decay to $\chi \bar \chi$, the increased optical depth from the abundant DM particles, manifested as $\Gamma_{\rm dC}$ in \Eq{L-A'}, dramatically increases the $A'$ optical depth and reduces the energy released in dark photons.

We will also study the two model variations mentioned in \Sec{subsec:models}, namely $\mathbf{(iii)}$~``{\bf inelastic dark matter}'' 
for several choices of the mass splitting $\Delta$, and $\mathbf{(iv)}$~``{\bf millicharged particles}''.  

We will not show any parameter space for $m_\chi \leq 100\kev$, but, as discussed at length above, these bounds do not decouple in 
the small $m_\chi$ limit. 
Note also that we can safely ignore the suppression due to the Landau-Pomeranchuk-Migdal effect, since this only suppresses production of particles with energy less than the quasiparticle width $\gamma$, with $\gamma \lesssim 5\mev$ for the densities of interest here \cite{Sedrakian:1999jh, Hanhart:2000ae}. In all cases, we will compare the 
SN1987A bounds to laboratory bounds and projections for proposed experiments, as appropriate.

\subsection{Heavy Dark Matter}\label{subsec:heavy-results}

For $m'<2m_\chi$, the energy is carried by both $A'$ and $\chi$ particles, since the $A'$ is stable at leading order in $\epsilon$. Production of $\chi$ particles is independent of $m_\chi$ as long as $m_\chi < \sqrt{3/2} \ \omega_p/2$, but is Boltzmann-suppressed at high masses. In contrast, $A'$ production is suppressed at small $m'$. For a given $\alpha_D$ not too small, the lower boundary on the $\ep$-constraint is thus determined from $\chi$ production when $m_\chi$ is small and from $A'$ production when $m_\chi$ is large. For large $\ep$, both $A'$ and $\chi$ particles are abundant, but the $A'$ can experience a large optical depth if a dense ``cloud'' of DM particles is created in the explosion,\footnote{These are different than the ``smog'' of relic DM particles that were suggested to affect the $A'$ optical depth in~\cite{Zhang:2014wra}.} written as $\Gamma_{\rm dC}(\omega,r')$ in \Eq{L-A'}. As discussed in \Sec{subsec:LA'}, to determine the upper boundary on the $\ep$-constraint,  
we use the more conservative bound (i.e.~the bound that excludes less parameter space) between the criterion from \Sec{subsec:trapping} and from~\cite{Chang:2016ntp}. 

\begin{figure}[t]
        \centering
            \includegraphics[width=0.495\textwidth]{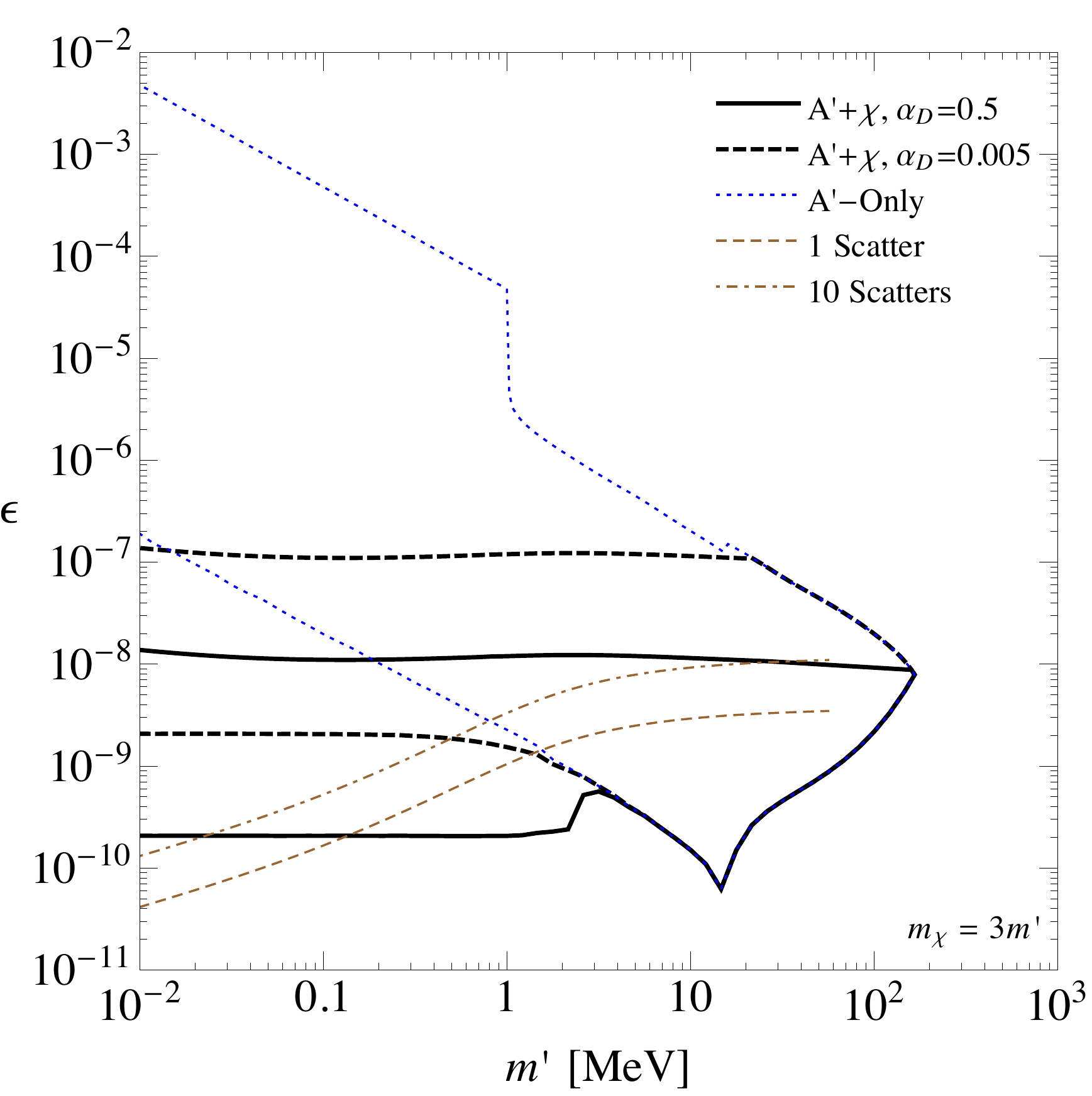}~~
            \includegraphics[width=0.495\textwidth]{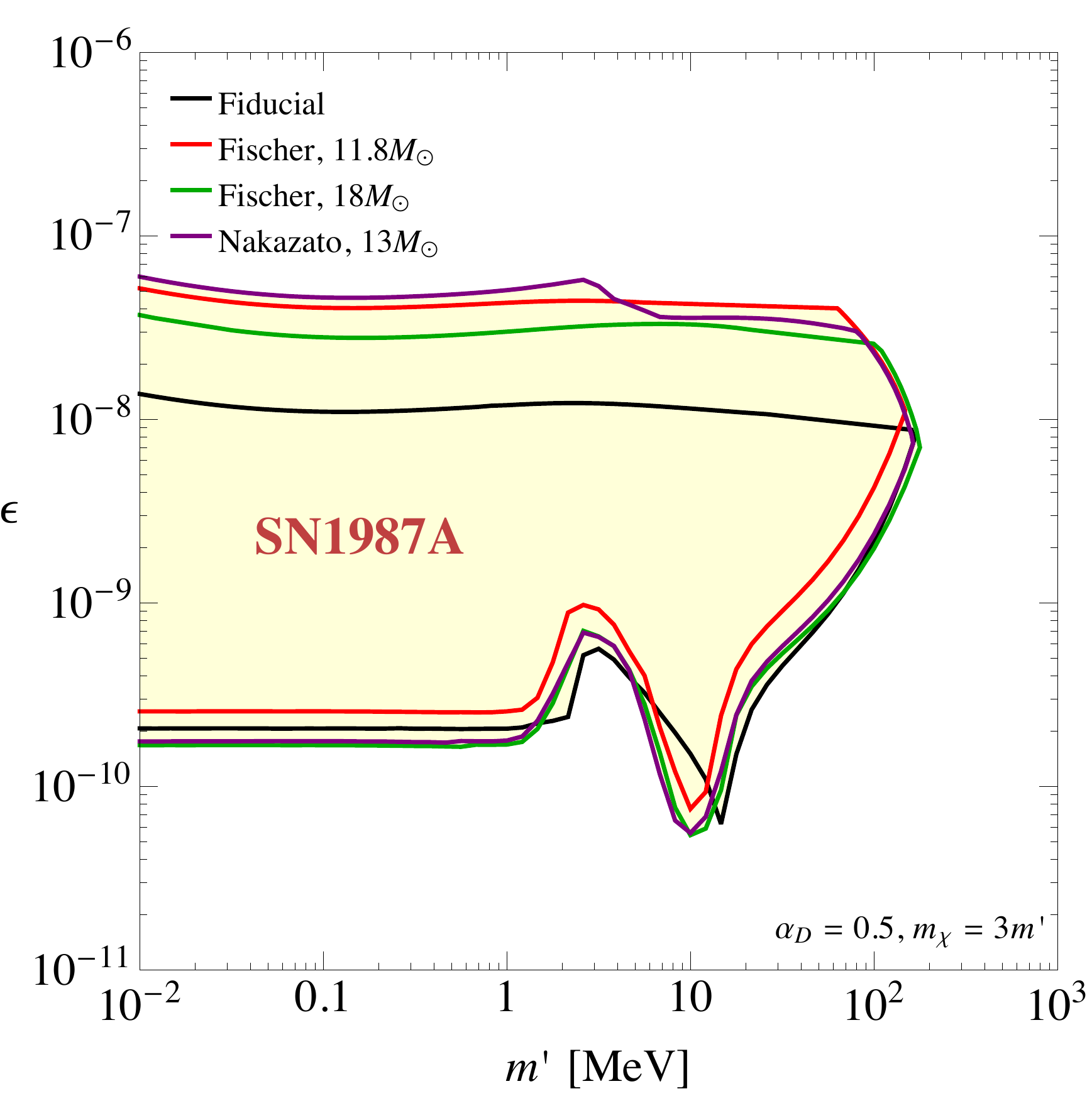}
        \caption[]
        {SN1987A constraints on ``heavy'' dark matter coupled to a dark photon, for the specific mass relation $m_\chi=3m'$.  
        {\bf Left:} Solid (dashed) black line shows the constraint for $\alpha_D=0.5$ ($\alpha_D=0.005$).  Along the brown dashed and dot-dashed lines, the dark matter scatters once and 10 times, respectively, on its way out of the star, for $\alpha_D=0.5$.   The blue dotted line is the constraint on a dark sector that contains only a dark photon and no dark matter.  We assume the fiducial temperature and density profile for the supernova.  
        {\bf Right:} Black lines are the same as in the left plot for the fiducial temperature and density profile, while colored lines are the constraints for the other profiles with $\alpha_D=0.5$.  
        }
        \label{heavydmplots1}
    \end{figure}

We show the resulting SN1987A bound in the left plot of \Fig{heavydmplots1}. The black solid and dashed lines are constraints on the heavy DM model for two values of $\alpha_D$, while the blue dotted line is the constraint for the $A'$-only case presented in~\cite{Chang:2016ntp}. The lower bounds are stronger for small $m_\chi$ than in the $A'$-only case because they are not lifted as $m' \to 0$, and they return to the same value as the $A'$-only case for $m' \simeq \omega_p$. For small $m_\chi$, as discussed above, the upper boundary lies below the $A'$-only case due to the large contribution of $\Gamma_{\rm dC}(\omega,r')$ to the optical depth. 

    \begin{figure}[t]
        \centering
            \includegraphics[width=0.485\textwidth]{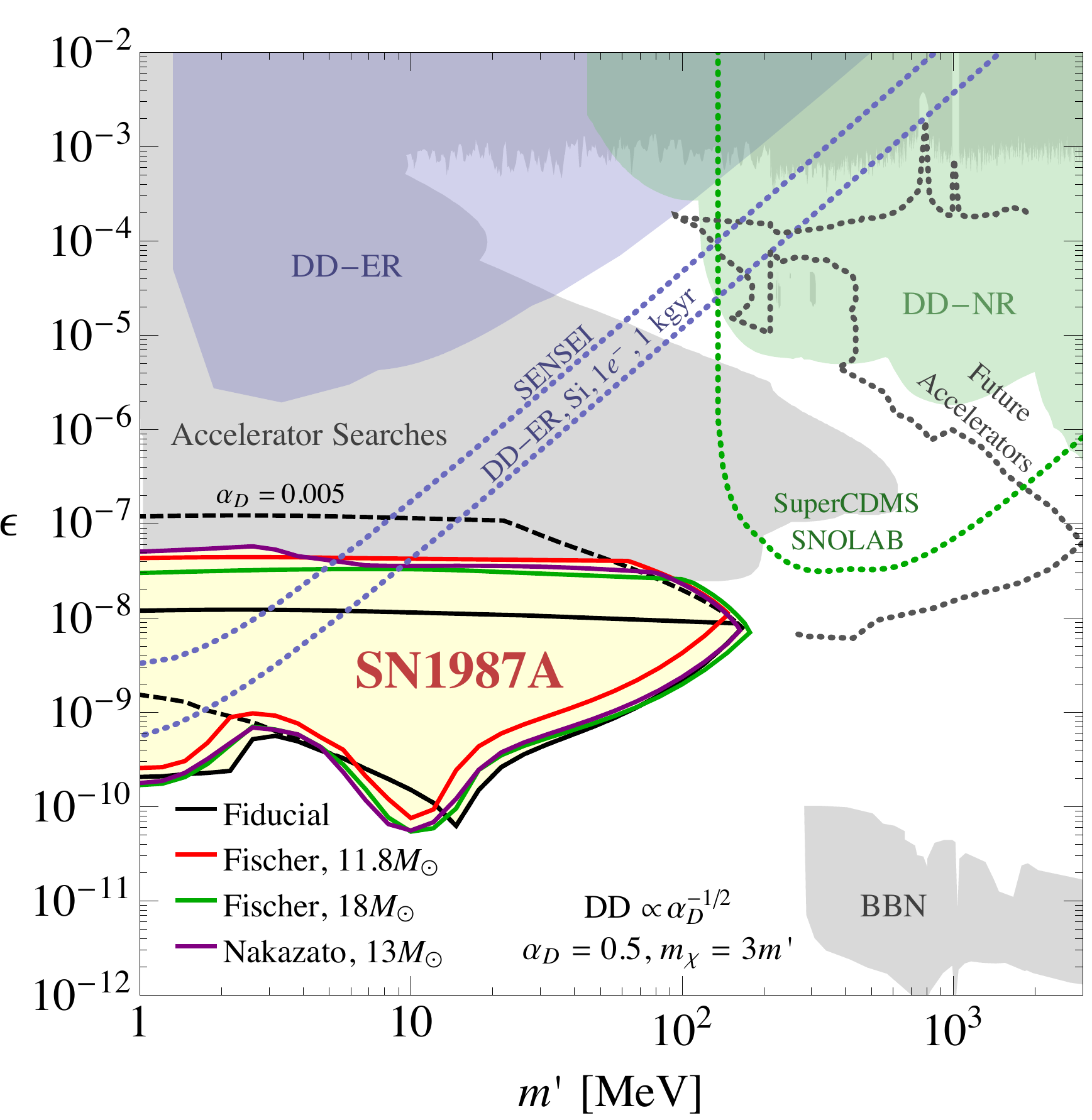}~~
            \includegraphics[width=0.51\textwidth]{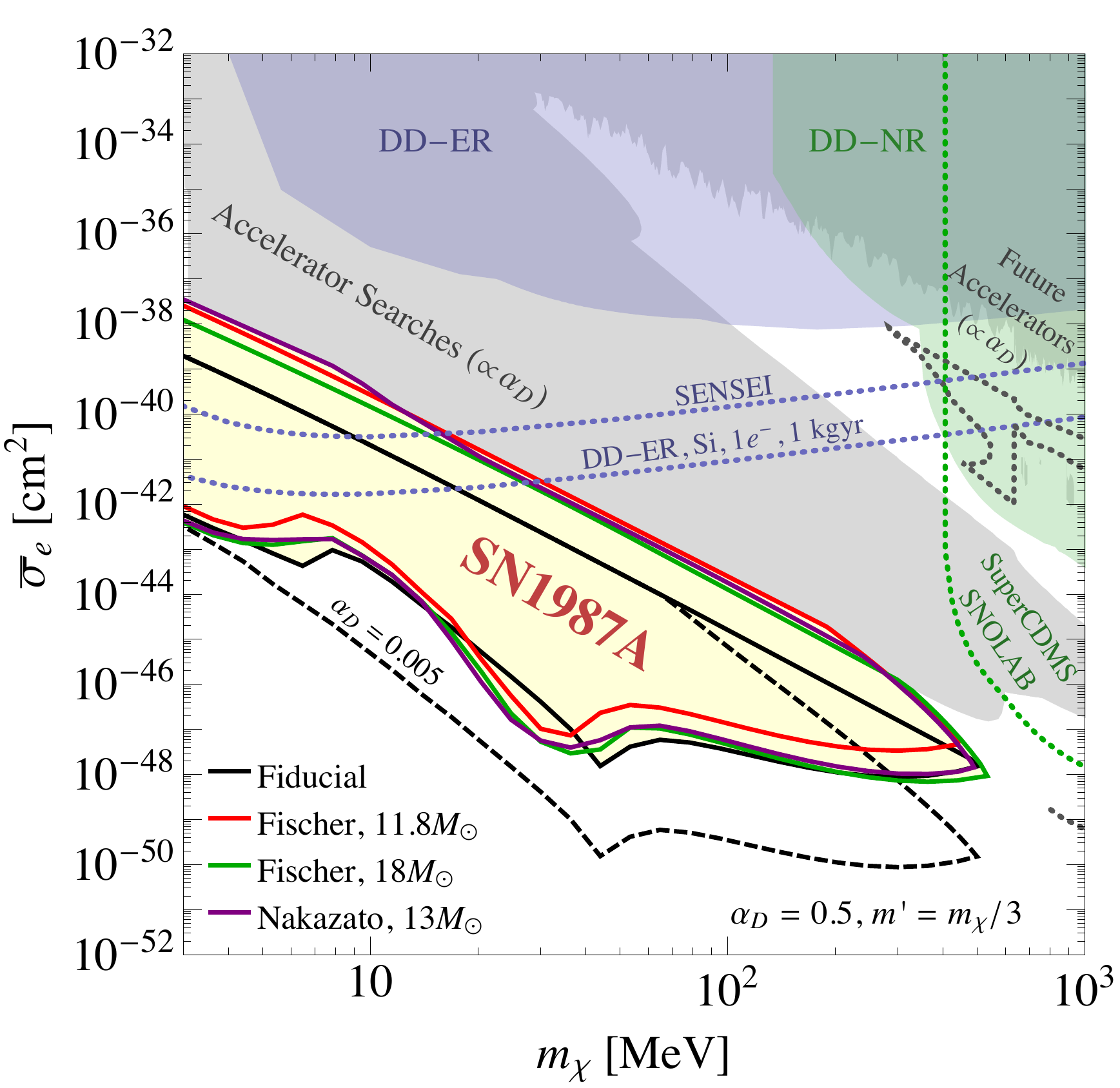}
        \caption[]
       { Thick solid black, red, green, and red lines show the SN1987A constraints on ``heavy'' dark matter coupled to a dark photon, 
       assuming the specific mass relation $m_\chi=3m'$, $\alpha_D=0.5$, and various temperature and density profiles.  
       Dashed black line shows the constraint for $\alpha_D=0.005$ using the fiducial profile.  
        {\bf Left:} The SN1987A constraints are displayed together with constraints from laboratory-based searches, including colliders, beam-dump 
and fixed-target experiments that search for $A'$ decays to Standard-Model particles.  
Under the assumption that the $\chi$ is all of the dark matter, we also show constraints on dark-matter electron scattering from 
XENON10, XENON100, and DarkSide-50, and constraints on dark-matter-nucleus scattering from the CRESST, SuperCDMS, and LUX 
collaborations.  
Dotted lines show projections from future collider and beam-dump searches (black), SuperCDMS SNOLAB (green), 
as well as SENSEI and a possible experiment using a silicon target sensitive to single electrons with a 1~kg-year exposure (both blue).  
The direct-detection constraints and projections scale as $\alpha_D^{-1/2}$.  
See text for references and details. 
{\bf Right:} Same as left plot, but in the $\sigma_e$ versus $m_\chi$ parameter space. 
        }
        \label{heavydmplots2}
    \end{figure}

For additional insight, we display the contour along which a typical $\chi$ scatters off a proton either once or ten times on the way out of the supernova with the brown dashed and dot-dashed lines, respectively. This diagnostic clearly gives us much less sensitivity than asking where the $\chi$ is expected to satisfy \Eq{random-walk}. This reflects a real physical effect: in order for a light DM particle coupled through a light mediator to become trapped and return to chemical equilibrium, it must scatter much more than once on its way out of the proto-neutron star. 

Varying $\alpha_D$ changes the asymptotically flat part of the upper boundary in $\ep$ such that $\alpha_D \ep^2$ is kept fixed, since this boundary is determined by the dark-matter-proton scattering cross section.  In addition, the flat part of the lower boundary in $\ep$ (i.e.~for small $m_\chi$) also changes such that $\alpha_D \ep^2$ is kept fixed, since that region is dominated by dark-matter pair production from bremsstrahlung (as opposed to $A'$ production with the $A'$ decaying to DM).  
In addition, while we do not show this explicitly, changing the mass ratio of $m_\chi/m'$ affects the value of $m_\chi$ below which the lower bound becomes independent of $m_\chi$. 

In the top-right panel of \Fig{heavydmplots1}, we show constraints for different temperature and density profiles as reviewed in \Sec{subsec:profiles} and given in~\cite{Chang:2016ntp}. The variation with different profiles can be taken as a systematic uncertainty on the bound.  
The upper boundary is higher for profiles that have a lower density beyond $R_\nu$ (see Fig.~3 in~\cite{Chang:2016ntp}), 
since it is easier in this case for the $\chi$ and $A'$ to leave the proto-neutron star.

It is interesting to show the constrained region in relation to laboratory searches for this dark sector model.  The left plot in 
\Fig{heavydmplots2} shows the SN1987A constraints together with the latest laboratory-based searches, including colliders, beam-dump 
and fixed-target experiments, and precision measurements~\cite{Bjorken:2009mm,Bjorken:1988as, riordan1987, bross1991, konaka1986,davier1989, andreas2012, Blumlein:1990ay, Blumlein:1991xh,Reece:2009un, Aubert:2009cp, Babusci:2012cr, Archilli:2011zc,Abrahamyan:2011gv, Merkel:2014avp,Agakishiev:2013fwl, Batley:2015lha,Battaglieri:2017aum,Aaij:2017rft,PhysRevLett.106.080801, Aoyama:2012wj, PhysRevLett.100.120801, Davoudiasl:2012ig}.  
The SN1987A bounds are complementary and constrain lower values of $\ep$ than these laboratory bounds.  
The plot also shows bounds from direct-detection experiments.  The bounds assume that the $\chi$ make up all the DM.  
There are two types of direct-detection bounds: from electron-recoil searches and from nuclear-recoil searches.  For the former, we use the constraints from~\cite{Essig:2017kqs,Essig:2012yx,Agnes:2018oej}, which are based on XENON10~\cite{Angle:2011th}, XENON100~\cite{Aprile:2016wwo}, and DarkSide-50 data \cite{Agnes:2018oej}, while for the latter, we use the combined bounds from the CRESST~\cite{Angloher:2017sxg,Petricca:2017zdp}, SuperCDMS~\cite{Agnese:2015nto}, and LUX~\cite{Akerib:2016vxi} collaborations.  
In order to put these bounds onto the $\ep$ versus $m'$ parameter space, we follow the definitions in~\cite{Essig:2011nj,Essig:2015cda} and define 
the direct-detection cross section for DM scattering off electrons (protons) as 
\beq
\bar\sigma_{e(p)} = \frac{16 \pi \alpha \alpha_D \ep^2}{m'^4} \, \mu_{\chi,e(p)}^2\,
\eeq
The electron-recoil searches constrain $\bar\sigma_e$, while the nuclear-recoil searches constrain $\bar\sigma_p$.  
Given a specific mass relation (we choose $m_\chi=3m'$) and value for $\alpha_D$ (we choose 0.5), we can display 
these constraints on the $\ep$ versus $m'$ parameter space.  
Note that for smaller values of $\alpha_D$, the direct-detection constraints will weaken as $1/\sqrt{\alpha_D}$, as indicated on the plot.  

We also show a combined projection from future collider and beam-dump searches (black dotted line)~\cite{Battaglieri:2017aum}, 
as well as from SuperCDMS SNOLAB~\cite{Agnese:2016cpb}, SENSEI~\cite{Essig:2015cda,Tiffenberg:2017aac}, and a possible search with an 
experiment using a silicon target sensitive to single electrons with a 1~kg-year exposure~\cite{Essig:2015cda}; for other projections, we 
refer the reader to~\cite{Battaglieri:2017aum}.  
These projections are largely complementary to the SN1987A bounds.  

One additional bound on this dark-sector model that we do not show on the plots 
can be relevant for large couplings and $m_\chi \lesssim 10$~MeV, assuming the $\chi$ are all of the DM and in 
thermal equilibrium with the SM sector in the early Universe.  
This bound comes from constraints on the effective number of degrees of freedom, $N_{\rm eff}$, 
from Big Bang Nucleosynthesis and the Cosmic Microwave Background~\cite{Boehm:2013jpa, Nollett:2013pwa}, since a light DM particle 
could affect the relation between the photon and neutrino temperatures after neutrino decoupling.  
Our SN1987A bounds are largely complimentary to this, since they apply to dark matter with small couplings.

\subsection{Light Dark Matter}\label{subsec:results-light}

    \begin{figure}[t]
        \centering
            \includegraphics[width=0.495\textwidth]{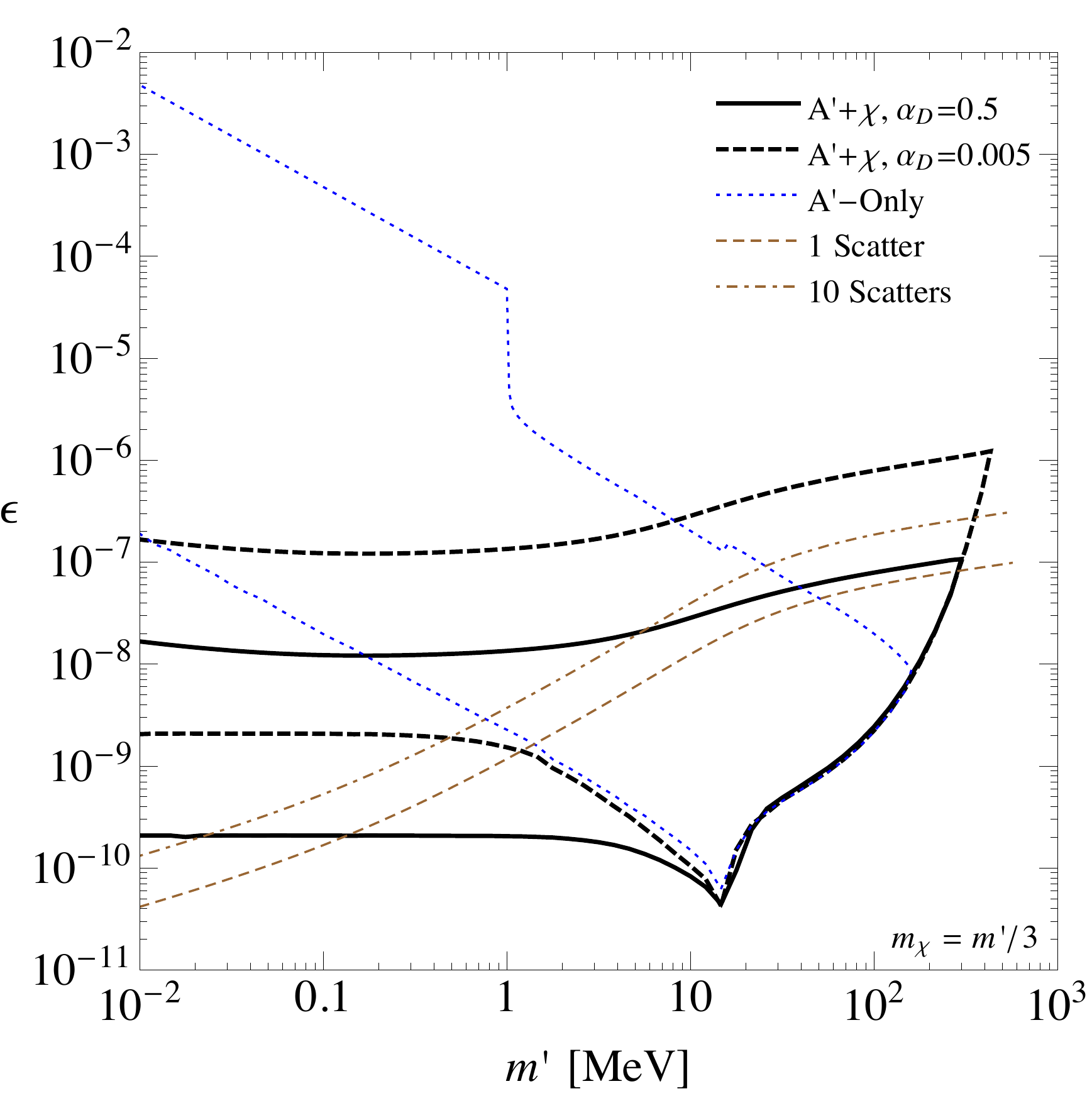}~~
            \includegraphics[width=0.495\textwidth]{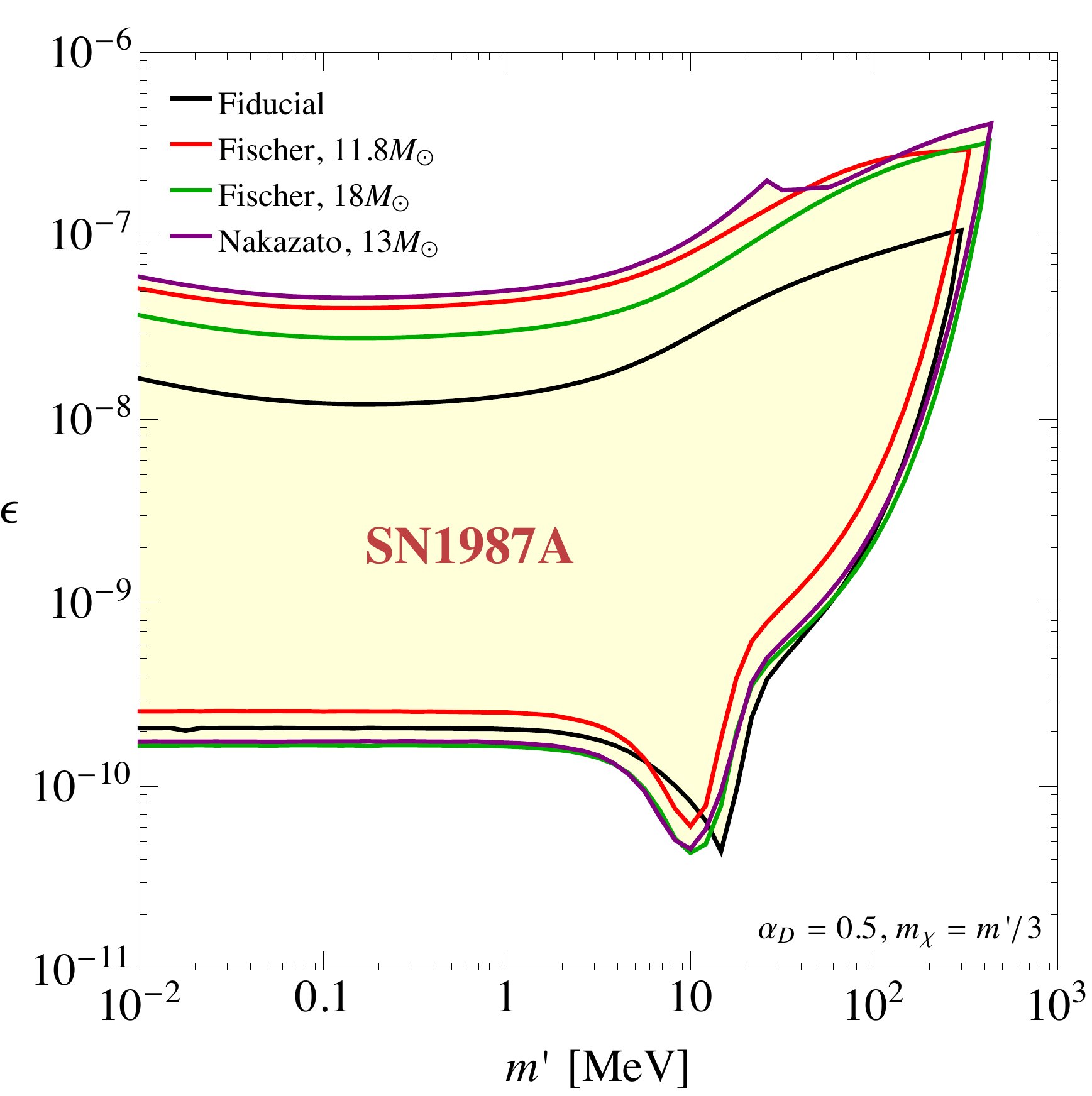} 
        \caption[]
        {SN1987A constraints on ``light'' dark matter coupled to a dark photon, for the specific mass relation $m'=3m_\chi$.  
        {\bf Left:} Solid (dashed) black line shows the constraint for $\alpha_D=0.5$ ($\alpha_D=0.005$).  Along the brown dashed and dot-dashed lines, the dark matter scatters once and 10 times, respectively, on its way out of the star, for $\alpha_D=0.5$.   The blue dotted line is the constraint on a dark sector that contains only a dark photon and no dark matter.  We assume the fiducial temperature and density profile for the supernova.  
        {\bf Right:} Black lines are the same as in left plot for the fiducial temperature and density profile, while colored lines are the constraints for the other profiles with $\alpha_D=0.5$.  
} 
        \label{lightdmplots}
    \end{figure}            
    
If the DM is ``light'', i.e.~for $2m_\chi<m'$, the dark photons will decay to DM pairs quickly (assuming $\alpha_D$ is not too small), 
so that all of the energy in the dark sector is in $\chi \bar \chi$ pairs. For this reason, the lower boundary of the SN1987A constraint is determined from \Eqs{diffP-DM}{SM-photon-decay-lum} and the upper boundary is determined from the criterion in \Eq{random-walk}.  

We show the resulting SN1987A constraint in \Fig{lightdmplots}.  
We use the same profiles and parameters as in \Fig{heavydmplots1}, except we now choose $m'=3m_\chi$ instead of $m'=m_\chi/3$.   
The qualitative features in both scenarios are largely the same, except that the upper boundary now increases at large $m'$ relative to \Fig{heavydmplots1}.  
This is because it is harder to trap the DM particles compared to the $A'$ alone, so that the decoupling radius $R_d$ decreases for 
$m_\chi \gtrsim T_c$, where $T_c \simeq 30 \mev$ is the supernova core temperature.

    \begin{figure}[t]
        \centering            
            \includegraphics[width=0.495\textwidth]{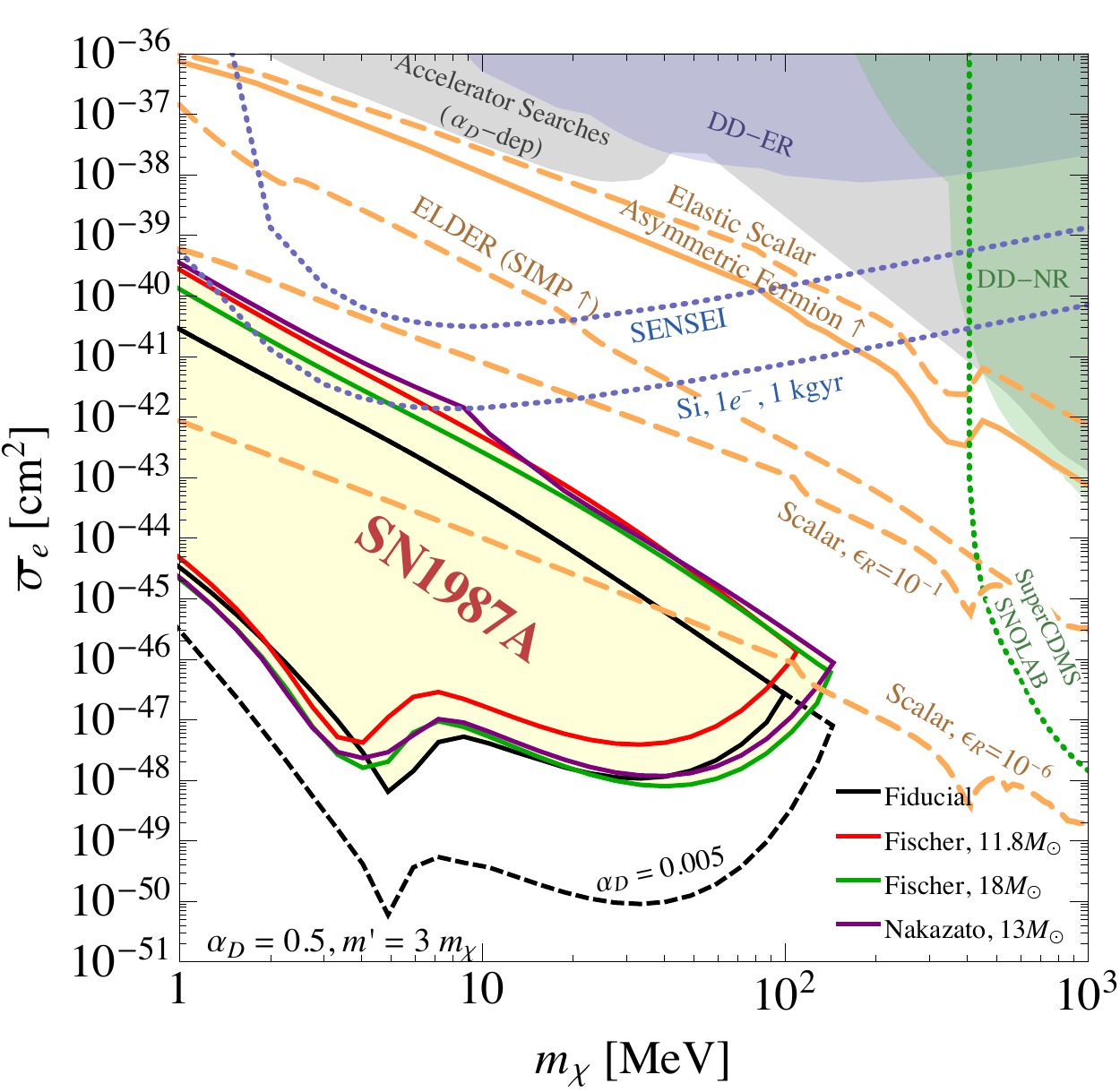}~~
            \includegraphics[width=0.495\textwidth]{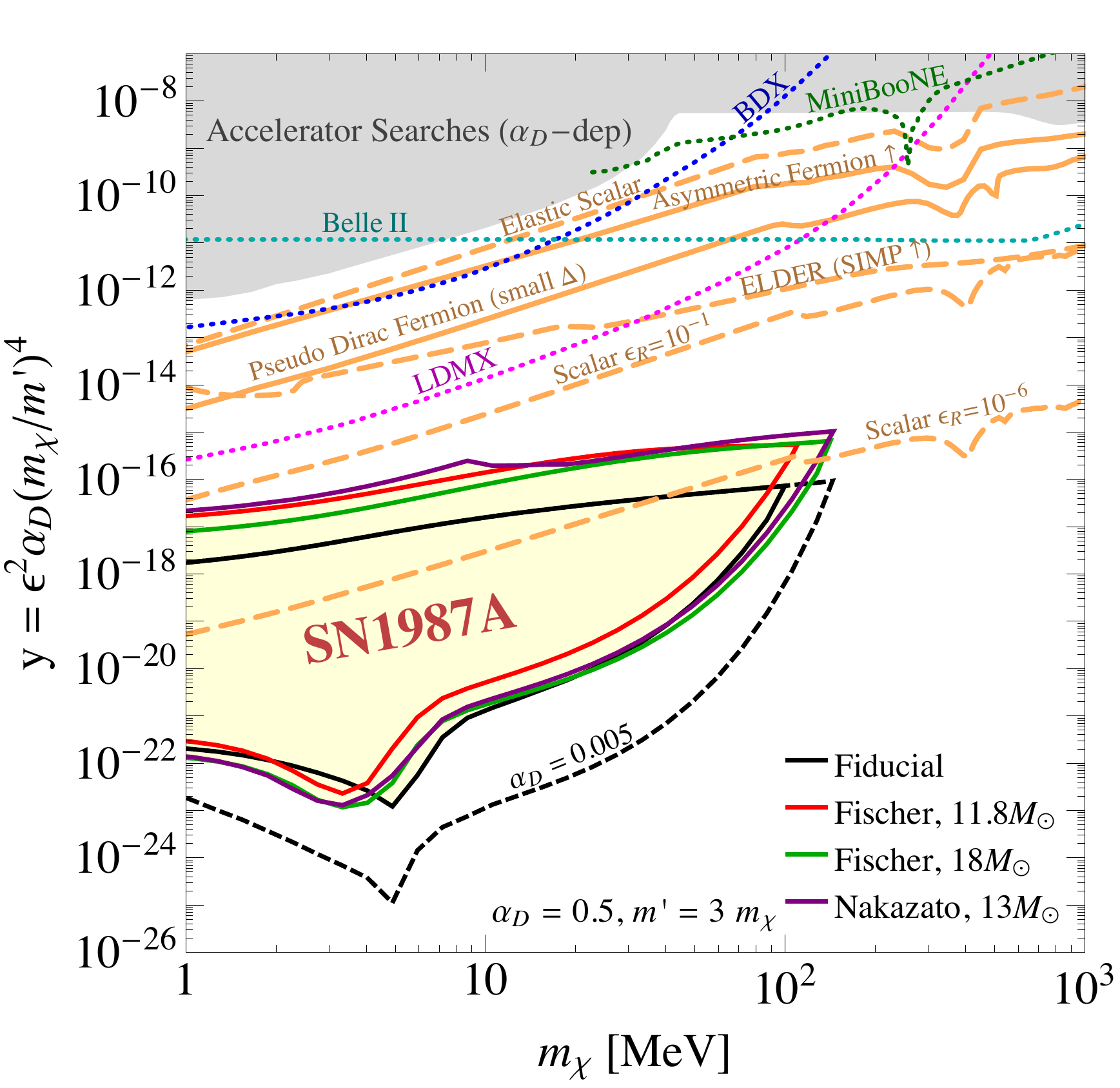}
        \caption[]
{ Thick solid black, red, green, and red lines show the SN1987A constraints for various temperature and density profiles 
on ``light'' dark matter coupled to a dark photon, 
       assuming the specific mass relation $m'=3m_\chi$, and $\alpha_D=0.5$.  
       Dashed black line shows the constraint for $\alpha_D=0.005$ using the fiducial profile.  Thick orange lines show several benchmark 
       model ``targets'', along (or above) which the DM can obtain the correct relic abundance in various scenarios.  
       Note that the SN1987A have been evaluated specifically for a DM particle that is a Dirac fermion (and not a scalar), 
while we show targets for both scalar and fermionic DM; however, the SN1987A constraints are expected to be similar in both cases (see discussion in \Sec{subsec:models}). 
       Existing laboratory-based searches are shown in gray, including colliders, beam-dump and fixed-target experiments that search for $A'\to\chi\bar\chi$ decay.  
        {\bf Left:} 
Under the assumption that the $\chi$ is all of the dark matter, we show constraints from dark-matter electron scattering from 
XENON10, XENON100, and DarkSide-50, and constraints on dark-matter-nucleus scattering from the CRESST, SuperCDMS, and LUX 
collaborations.  
Dotted lines show projections SuperCDMS SNOLAB (green), as well as SENSEI and a possible experiment using a silicon target sensitive to single electrons with a 1~kg-year exposure (both blue).  
{\bf Right:} Same as left plot, but in the $y$ versus $m_\chi$ parameter space.  We again show the same accelerator-based constraints as in the left plot, but now show projections in dotted lines from Belle-2 (cyan) as well as the proposed experiments BDX (blue), LDMX (magenta), and MiniBooNE (dark green). 
See text for references and details.
} 
        \label{lightdmconstplots}
    \end{figure}

It is again instructive to show the SN1987A constraints for this particular dark-sector model together with other, laboratory-based constraints and projected sensitivities from selected future direct-detection and accelerator-based experiments.  
We show this in \Fig{lightdmconstplots}, in the $\sigma_e$ versus $m_\chi$ plane (left) and the $y$ versus $m_\chi$ plane (right), where 
\beq
y \equiv \alpha_D \ep^2 \frac{m_\chi^4}{m'^4}\,.
\eeq
The accelerator constraints are based on LSND~\cite{Batell:2009di}, E137~\cite{Batell:2014mga}, 
BaBar~\cite{Essig:2013vha,Lees:2017lec}, and MiniBooNE~\cite{Aguilar-Arevalo:2017mqx}, and are discussed in 
detail in e.g.~\cite{Izaguirre:2015yja,Essig:2015cda}.  Projections for accelerator-based searches are shown for Belle-2~\cite{Essig:2013vha}, 
BDX~\cite{Izaguirre:2014dua,Battaglieri:2016ggd}, LDMX~\cite{Izaguirre:2014bca}, and MiniBooNE~\cite{Battaglieri:2017aum}.  
The direct-detection constraints and projections are the same as in \Sec{subsec:heavy-results}.  
For other projections see~\cite{Battaglieri:2017aum}.  
The bounds and projections from accelerator-based searches strengthen for smaller $\alpha_D$: 
for experiments in which the DM is produced in a beam dump and then scatters in a downstream detector 
(LSND, E137, BDX, and MiniBooNE), the bound scales as $\sqrt{\alpha_D}$, while for experiments searching for missing-energy signals 
(BaBar, Belle-2, LDMX), 
the bound scales as $\alpha_D$.
The direct-detection constraints and projections do not change when varying $\alpha_D$.  
We see that a significant amount of parameter space is unconstrained by the SN1987A bound, 
and ripe for exploration by these future searches.

The thick orange lines show specific experimental ``targets'' corresponding to several benchmark models 
(they do not change when varying $\alpha_D$). 
These are based on work from~\cite{Boehm:2003hm,Izaguirre:2015yja,Essig:2015cda,Hochberg:2014dra,Kuflik:2015isi,Kuflik:2017iqs,Feng:2017drg}, and we refer the reader to~\cite{Battaglieri:2017aum} for a summary.  
Note that we derived the SN1987A constraints for a dark sector consisting only of a Dirac fermion that is coupled to a dark photon (solid lines). 
Some of the targets shown (in dashed) assume a scalar DM particle, additional interactions within the dark sector, and/or 
a resonance in the process $\chi+\chi \to A'^* \to {\rm SM+SM}$. A scalar $\phi$, even with strong self-interactions (as a SIMP), 
will have a similar production rate as fermions, and the upper bound will only change by an equivalent number of effective
blackbody degrees of freedom, $\sim g_\phi/(g_\chi\times 7/8)$.
Likewise, a resonance in the DM annihilation cross section, parameterized by $\ep_R\equiv (m'^2-4m_\chi^2)/4m_\chi^2$, lowers the required couplings
to achieve the correct relic abundance, but this resonance does not impact the dark-sector production rate in the proto-neutron star 
to an appreciable extent.
None of these changes to the dark sector content will thus drastically affect the SN1987A constraint, and we find it instructive 
to show all the ``targets'' on the same plot. 
We see that most of the targets are unconstrained by the SN1987A bound; only the resonant thermal targets with $\ep_R\lesssim 0.1$ are 
partly constrained.

\subsection{Inelastic Dark Matter}
\label{subsec:results-iDM}

We now discuss the SN1987A constraints on an inelastic DM model consisting of two states, $\chi_1$ and 
$\chi_2$.  
We will only consider the ``light'' DM scenario, where the dark photon is heavy and allows for the decay $A'\to \chi_1\chi_2$. 
As discussed in \Sec{subsubsec:iDM-model}, we will focus on the case where the elastic, tree-level coupling $\chi_i\chi_i$ 
($i=1,2$) vanishes. If such a coupling is present, it is velocity suppressed.  We thus expect the bounds at small DM masses to be 
similar to the elastic case discussed in previous sections, but at large DM masses, $\gtrsim\! T_c$, when the DM 
does not have much kinetic energy, the bound will likely be similar to the inelastic case discuss in this subsection.
Defining $\Delta\equiv m_2-m_1$, there are two cases of interest: (i) $\Delta \ll m_1\simeq m_2$ and (ii) $\Delta \simeq m_1$.  
For case (i), the bounds are essentially the same as the elastic cases discussed in the previous sections.  
However, for larger $\Delta$, i.e.~case (ii), the SN1987A bounds become significantly stronger at large couplings (along the upper boundary) 
for $\chi_1$ with masses above $T_c$, since it is harder for the DM particles to scatter and become trapped.  

    \begin{figure}[t]
        \centering
            \includegraphics[width=0.495\textwidth]{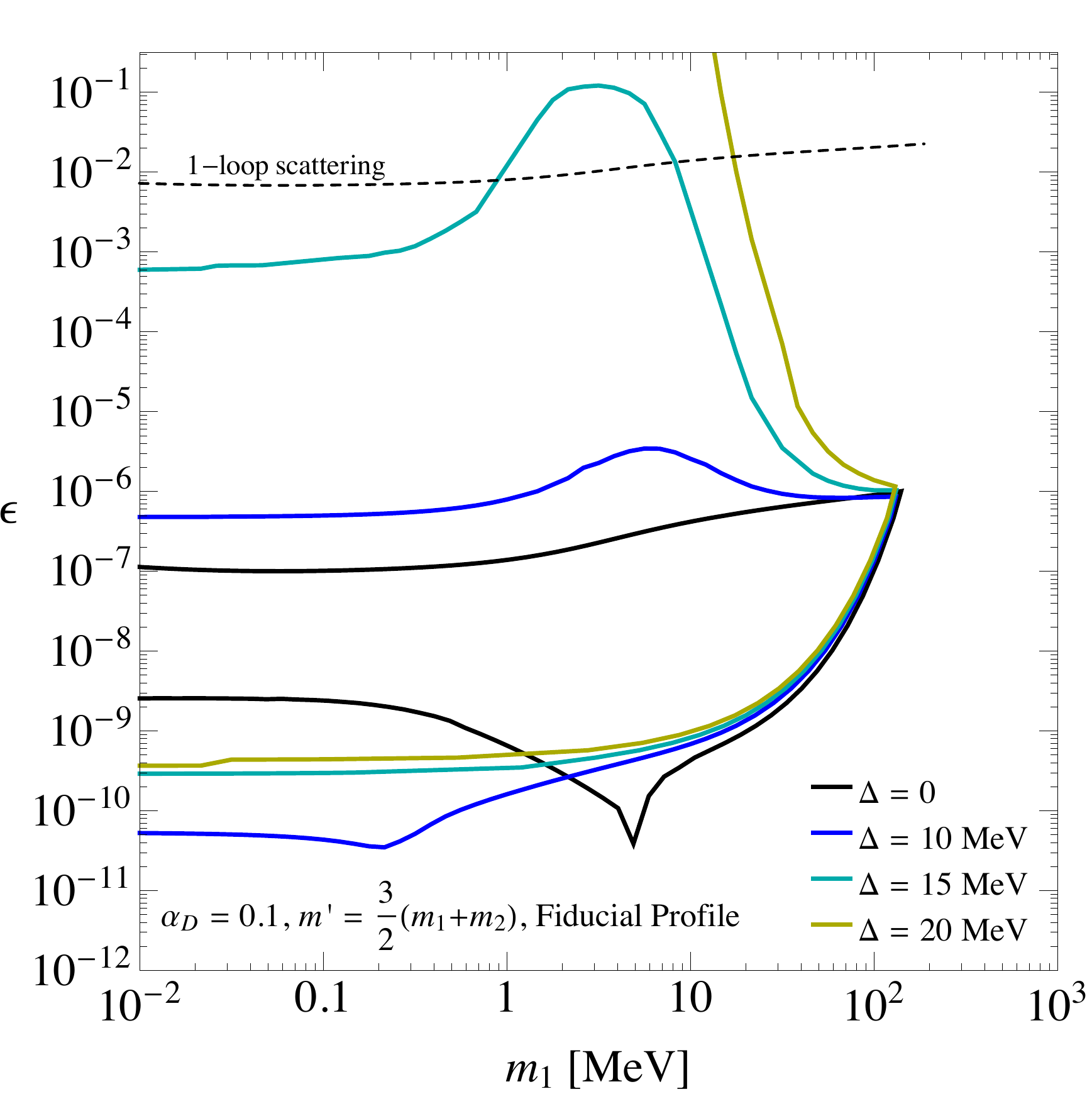}~~
            \includegraphics[width=0.495\textwidth]{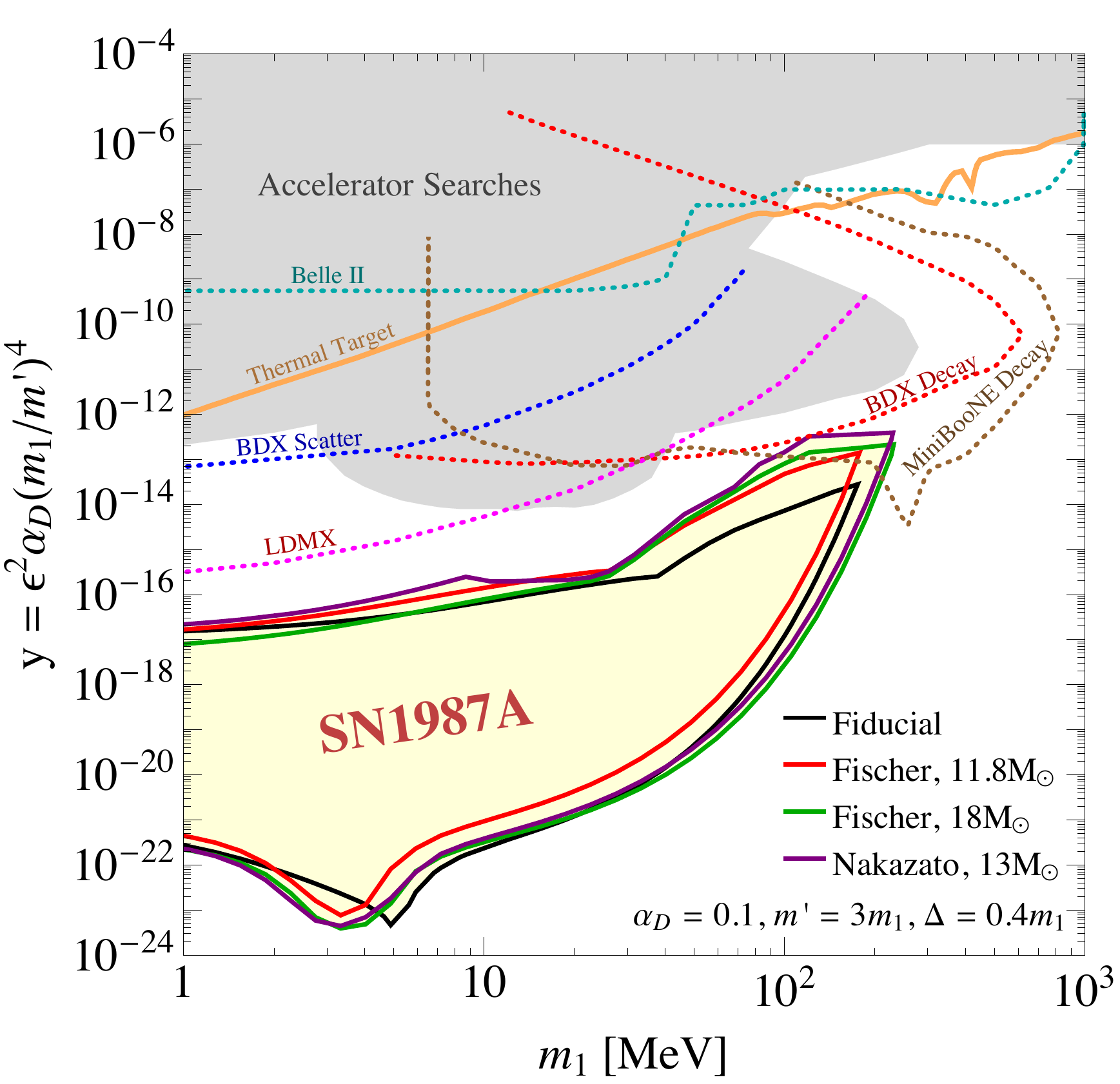}
        \caption[]
{{\bf Left:} Solid colored lines show the SN1987A constraints on inelastic dark matter consisting of two states $\chi_1$, $\chi_2$ 
with various mass splittings $\Delta$, where $\Delta\equiv m_2-m_1$.  We use the fiducial temperature and density profile, and 
set $\alpha_D=0.1$ and $m'=\frac{3}2(m_1+m_2)$.   The solid black line shows the elastic case $\Delta=0$.  
The dotted line shows the approximate value of $\ep$ above which the $\chi_1$ is trapped by the two-dark-photon-exchange 
process at one-loop allowing for an elastic scatter of $\chi_1$ to $\chi_1$.  
{\bf Right:} Thick solid black, red, green, and red lines show the SN1987A constraints in the $y$ versus $m_1$ parameter space 
for various temperature and density profiles on inelastic dark matter with $\Delta = 0.4 m_1$, $\alpha_D=0.1$, and $m'=3m_1$.  
Existing laboratory-based searches are shown in gray, including colliders, beam-dump and fixed-target experiments, and 
projections from proposed experiments are shown in colored dotted lines~\cite{Izaguirre:2017bqb}. 
} 
        \label{idmplots}
    \end{figure}

Let us discuss case (ii) in more detail.  Here $\chi_1$ and $\chi_2$ are produced from (on-shell) $A'$ decay and via bremsstrahlung in 
proton-neutron collisions.   
However, for sizable $\Delta$, any $\chi_2$ that is produced will quickly decay to $\chi_1 e^+e^-$ through an on- or off-shell $A'$, so that 
the proto-neutron star essentially contains only $\chi_1$. 
In order for the $\chi_1$ to become trapped, they must scatter off protons into the heavier particle $\chi_2$.   
This is only possible for those $\chi_1$ that find a proton with energy $\gtrsim \Delta $;  the 
population of such protons is exponentially suppressed if $\Delta \gtrsim T_c$.  
Therefore, if $\Delta \gtrsim 15 \mev$, even very large couplings will be excluded by the SN1987A data, since the $\chi_1$ can freely 
escape.  

A simulation is required to calculate the upper boundary accurately for large $\Delta$: the $\chi_1$ scatter into $\chi_2$, which in turn 
decay to $\chi_1 e^+e^-$, with the resulting $\chi_1$ typically having less energy than the original $\chi_1$.  This process can then 
repeat multiple times as the $\chi_1$ attempt to escape the proto-neutron star.  
It is computationally challenging to calculate the upper boundary using our trapping criterion, \Eq{random-walk}, as we did 
for the elastic case. Instead, we will use a simpler and very conservative criterion: 
we calculate the couplings needed for which a typical $\chi_1$ scatters off a proton {\it once} on its way out of the proto-neutron star.  
This criterion is appropriate given the other uncertainties and also because after a single scatter a good fraction of the energy is immediately 
reprocessed into the SM sector via the $e^+e^-$ produced in the $\chi_2$ decay.  

We present our results in \Fig{idmplots}.  The left plot shows the constraint in the $\ep$ versus $m_{\chi_1}$ plane for various $\Delta$, 
for the fiducial temperature and density profiles and $\alpha_D=0.1$.  
The upper boundary for the $\Delta=0$ constraint uses our trapping criterion, \Eq{random-walk}, while the upper boundaries 
for $\Delta\ne 0$ are derived by requiring the $\chi_1$ to scatter once as discussed above. 
As expected, the upper boundary of the bounds strengthens dramatically for $m_1 \ll m_2$.  

The SN1987A data constrains very large couplings for large $\Delta$.  However, for very large couplings, a two-dark-photon-exchange 
process at one-loop allows for an elastic scatter of $\chi_1$ to $\chi_1$, which can dominate over the kinematically suppressed 
$\chi_1\to \chi_2$ transition.  
We do not calculate in detail this one-loop diagram, but give a simple estimate above which the bounds shown in 
solid lines in \Fig{idmplots} are not applicable.  
The cross section for the one-loop diagram is proportional to $\alpha^2 \alpha_D^2 \ep^4/16\pi^2$, while for 
the tree-level $A'$-exchange process, the cross section scales as $\alpha \alpha_D \ep^2$.  
In order to estimate when the one-loop elastic process is important in trapping the $\chi_1$, we simply set 
\beq
\alpha \alpha_D \ep^2 |_{\rm tree-level} = \frac{\alpha^2 \alpha_D^2 \ep^4}{16\pi^2} \rvert_{\rm one-loop} \,. 
\eeq
The left-hand side of this equation is set by the value of $\alpha_D\ep^2$ calculated for $\Delta=0$ (the elastic case) 
with our trapping criterion, \Eq{idmplots}; setting this equal to $\alpha_D\ep^2$ on the right-hand side then determines when 
the elastic one-loop scattering process contributes at a similar level. 
We find that $\ep\simeq 7\times 10^{-3}$ for $\alpha_D=0.1$, which is indicated by the dotted line in \Fig{idmplots} (left).

The right plot in \Fig{idmplots}, shows the constraints on the $y$ versus $m_{\chi_1}$ parameter space for $\Delta = 0.4 m_{\chi_1}$ for 
our four temperature and density profiles.  
Here the upper boundary of the SN1987A bound is derived by requiring either the trapping criterion for $\Delta=0$ or a single scatter, 
whichever is stronger.  
We see that for $\Delta\gtrsim T_c$, the SN1987A data constrains larger couplings, while for smaller $\Delta$, the upper boundary is essentially 
the same as in the elastic case.  
We also show current constraints from accelerator-based searches (in gray) and projections from proposed experiments (dotted lines), 
including Belle-2, MiniBooNE, BDX, and LDMX~\cite{Izaguirre:2017bqb}.   
We again see that existing constraints, projected searches, and the SN1987A constraints are all complimentary and probe different 
regions in parameter space.  

\subsection{Millicharged Particles}
\label{millichargesection}

In this subsection, we consider millicharged particles, as discussed in \Sec{subsubsec:milli-model}.  
Our SN1987A constraints improve on prior work~\cite{Mohapatra:1990vq,Davidson:2000hf} by considering the plasma effects on the SM photon, 
an improved trapping criterion, and an improved treatment of the high-mass region. 
We also consider several detailed temperature and density profiles. 

\begin{figure}[t]
        \centering
            \includegraphics[width=0.495\textwidth]{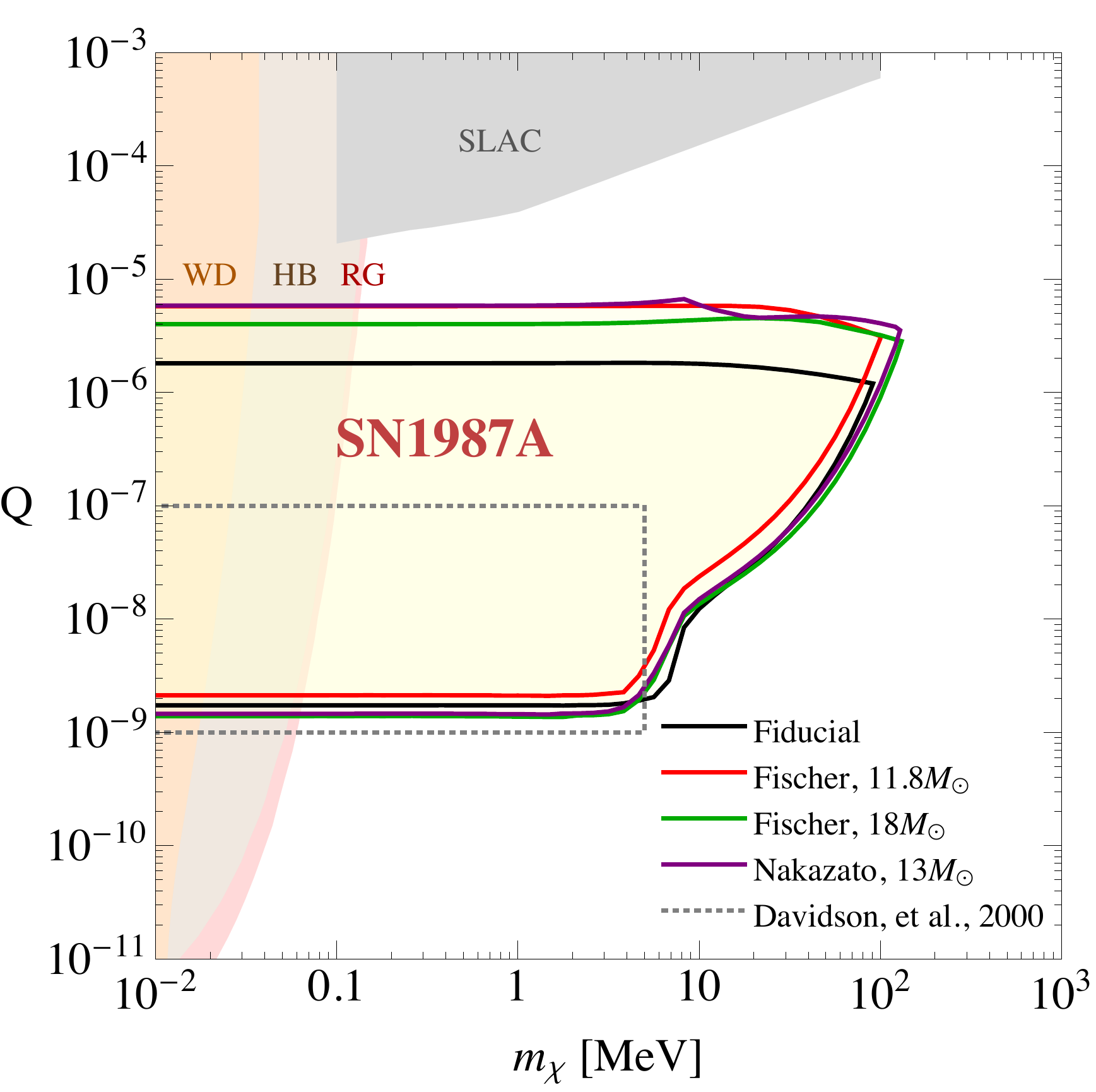}~~
            \includegraphics[width=0.495\textwidth]{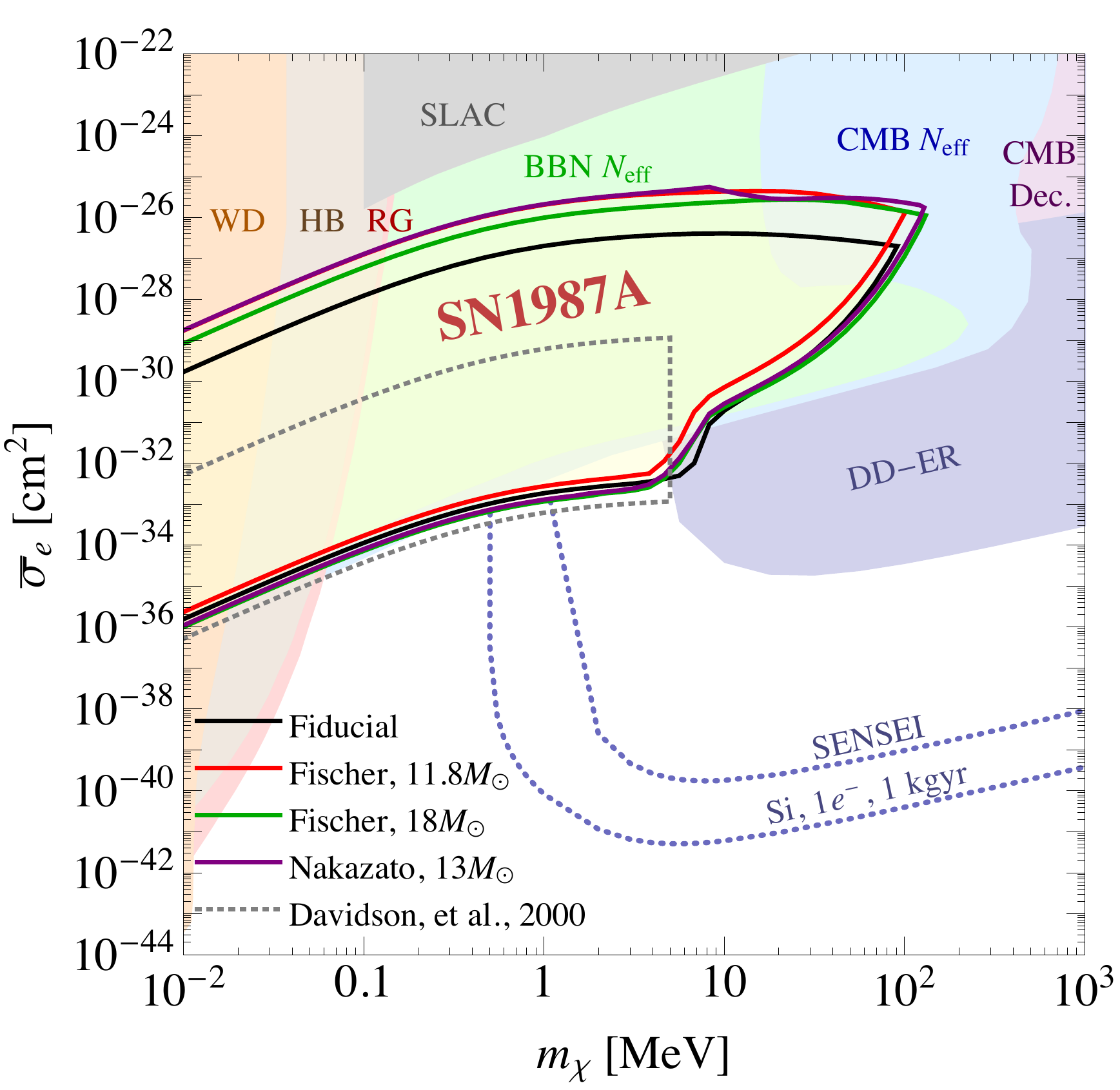}
        \caption[]
{
{\bf Left:} Thick solid black, red, green, and red lines show the SN1987A constraints for various temperature and density profiles 
on millicharged particles, updating the bounds presented in~\cite{Davidson:2000hf} (dotted line).  
Other constraints on millicharged particles are taken from~\cite{Vogel:2013raa,Prinz:1998ua,McDermott:2010pa}.
{\bf Right:} The constraints on millicharged particles are also applicable to dark matter coupled to an ultralight dark-photon mediator, which 
we show here in the $\bar\sigma_e$ versus $m_\chi$ parameter space.  
We also show a bound on dark-matter-electron scattering using XENON10 data (shaded light blue region)~\cite{Essig:2012yx}, 
and projections from the upcoming direct-detection experiment SENSEI and a possible experiment using a silicon target sensitive to single electrons 
with a 1~kg-year exposure (both dotted blue)~\cite{Essig:2015cda,Tiffenberg:2017aac}.
} 
        \label{millichargeddmplots}
    \end{figure}

Our main results are shown in \Fig{millichargeddmplots} (left) in the $Q$ versus $m_\chi$ parameter space.  The solid colored lines show 
the constraint from using different temperature and density profiles for the proto-neutron star.  
Note that plasma effects self-consistently cut off the potential divergence at low-momentum transfers in our calculation.  
The dotted line shows the constraint from~\cite{Davidson:2000hf}.  While our lower boundary is slightly higher, our upper boundary is stronger by more than an order of magnitude. 
We also show constraints from the SLAC millicharge experiment~\cite{Prinz:1998ua}, as well as white-dwarf, red-giant, and horizontal-branch stars, all of which are independent of whether the $\chi$ is present in the early Universe~\cite{Vogel:2013raa}. In addition, we show constraints from $N_{\rm eff}$ considerations at the time of BBN and the CMB, assuming no dark sector population after reheating~\cite{Vogel:2013raa}, 
and we also show a region in which the DM has not decoupled from the SM at the time of the formation of the CMB~\cite{McDermott:2010pa,Xu:2018efh}.

The constraints on millicharged particles can also be applied to DM interacting with a massive, but ultralight, mediator.  Such a mediator 
can mediate DM-electron scattering, leading to a cross section that scales as $1/q^4$, where $q$ is the momentum transfer.  
We show the SN1987A constraints on the $\bar\sigma_e$ versus $m_\chi$ plane in \Fig{millichargeddmplots} (right), together again 
with the other constraints also shown in the left plot.  We also now include a constraint on DM-electron scattering 
using XENON10 data from~\cite{Essig:2012yx}.  This plot updates the bounds presented in~\cite{Essig:2015cda}.  
Projections from selected future direct-detection experiments are shown with dotted lines~\cite{Essig:2015cda,Tiffenberg:2017aac}; 
for other projections see~\cite{Battaglieri:2017aum}. 

Note that since our upper boundary is more than an order of magnitude stronger than prior bounds, it could have ramifications for the
recently claimed detection of an anomalous absorption strength in the 21cm line from the epoch of first star formation~\cite{Bowman:2018aa, Barkana:2018aa}; for example, our bounds disfavor some of the parameter space shown to be open in~\cite{Munoz:2018pzp}.

\section{The Hadronic QCD Axion}
\label{axionsection}

We now shift to discuss a different DM candidate, the QCD axion~\cite{Preskill:1982cy, Abbott:1982af, Dine:1982ah}. Unlike the fermionic DM discussed in \Sec{sec:DM}, the axion has no conserved quantum numbers. This enables us to make a straightforward calculation of the luminosity as in \Eq{L-A'}, where we must of course apply suitable substitutions for the bremsstrahlung production rate and the optical depth. In this section, we evaluate this luminosity for the KSVZ or ``hadronic'' axion. 
The KSVZ axion couples to the CP-odd combination of gluon and hypercharge field strengths, and also to nucleons with 
\bea
\cL \supset \sum_N \frac{C_N}{2f_a} \partial_\mu a \bar N \gamma^\mu \gamma_5 N\,. 
\eea
On the equations of motion, the fermion mass may be substituted for the derivative, $\partial_\mu \bar f \gamma^\mu \gamma_5 f \to 2i m_f \bar f \gamma_5 f$, such that $\cL \supset -i \sum_N \frac{m_N C_N}{f_a} \partial_\mu a \bar N \gamma^\mu \gamma_5 N$, where $m_N$ is the mass of a nucleon.

Calculations for the axion bremsstrahlung rate have been obtained previously under a variety of simplifying assumptions. Limits on the axion coupling and mass have been extracted in these contexts starting immediately after the observation of SN1987A. We provide a chronological summary of related prior work in \App{previous-axion}. Here, we evaluate the limit on the axion beyond the diagrammatic calculation of the nuclear scattering cross section and without approximating the luminosity as a blackbody spectrum at large coupling, where absorption is important. Instead, we use results for the spin-flip current obtained at N${}^3$LO order in chiral perturbation theory~\cite{Bacca:2008yr, Bartl:2014hoa, Bartl:2016iok} to consistently ``correct'' the diagrammatic rate. The higher order contributions are stable, but due to a large, well-understood destructive interference at NLO they display qualitative differences from the leading order result. In order to make the comparison with previous bounds as clear as possible, we will phrase the N${}^3$LO results in terms of multiplicative corrections to the leading order result. In practice, we multiply the tree-level result by suitable density- and energy-dependent factors to reproduce the N${}^3$LO result.

These corrections change existing limits in important ways. 
At low (high) coupling, our constraints point to a bound on the axion mass that is a factor of about five (one to two orders of magnitude) higher 
than previously extracted~\cite{Patrignani:2016xqp}. 
Equivalently, this implies a bound on the Peccei-Quinn breaking scale that is lower by a factor of about five (one to two orders of magnitude).  
We discuss the nature of these corrections now.

\subsection{Corrections to the Axion Bremsstrahlung Rate}
\label{axion-corrections}

Our results incorporate three classes of corrections to the tree-level, massless pion calculation: 
a cutoff for scattering at arbitrarily low energies, 
a factor for the nucleon phase space that accounts for the finite pion mass, and a factor that introduces higher orders in the nucleon scattering. These effects have been known in some cases for many years, but they have not been consistently applied to the scattering rate of the axion.

\begin{figure}[t]
\begin{center}
\includegraphics[width=0.95\textwidth]{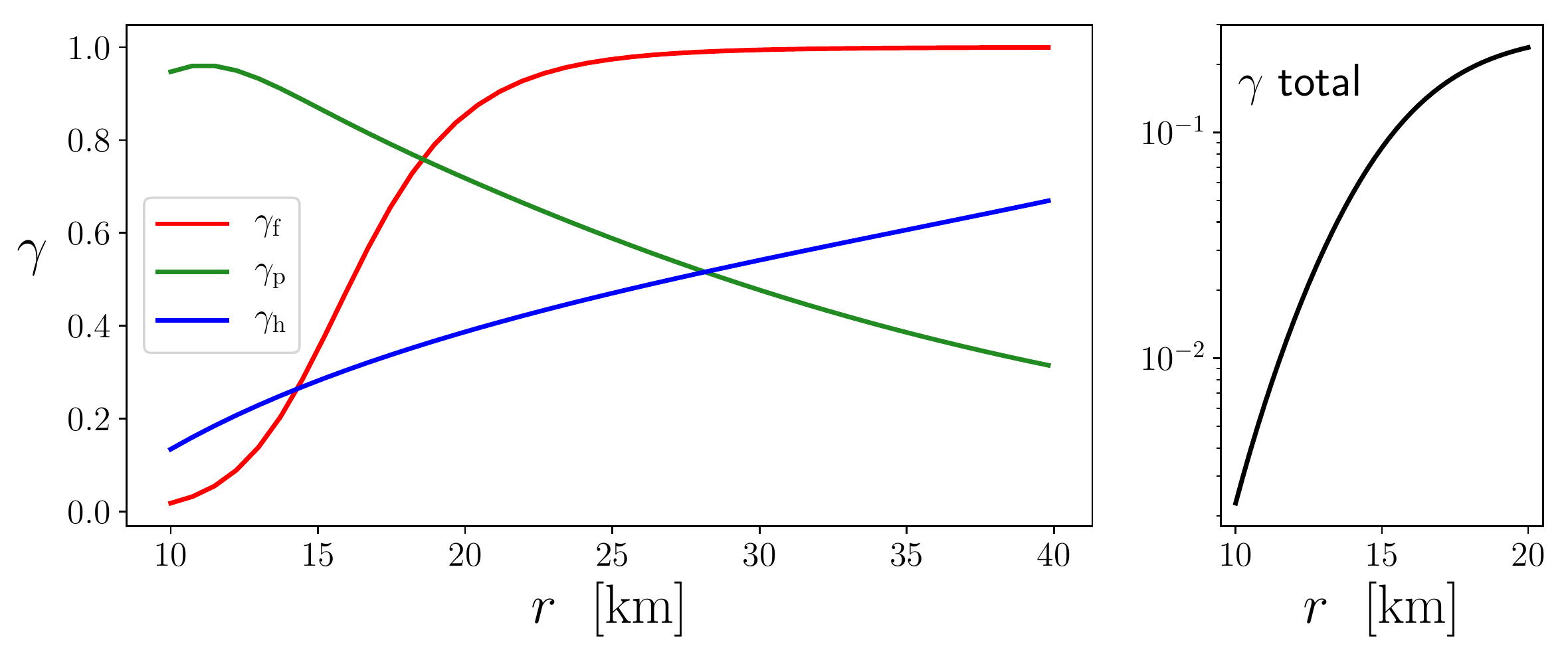}
\caption{Each of the correction factors $\gamma$ shown in the {\bf left} panel multiplies the total rate, as in \Eq{corrected-rate}. For radii close to the core the suppression is more than two orders of magnitude, so we zoom in on the product of corrections at small radii in the {\bf right} panel.}
\label{corrections}
\end{center}
\end{figure}

Collecting all effects and setting the notation, we write an amended form of the canonical expression for the axion absorptive width (found, \eg, in~\cite{Raffelt:2006cw}) as
\beq \label{corrected-rate}
\Gamma_a= \Gamma_a^{nn}+ \Gamma_a^{pp} + \Gamma_a^{np} + \Gamma_a^{pn}, \qquad \Gamma_a^{ij} = \frac{C_i^2 Y_i Y_j}{4f_a^2}\frac{\omega}2 \frac{n_B^2 \sigma_{np\pi}}{\omega^2} \gamma_{\rm f} \gamma_{\rm p} \gamma_{\rm h}\,.
\eeq
The factors that appear in \Eq{corrected-rate} are:
\begin{itemize}
\item[$C_i$] is the coupling of the axion to nucleon  $i=n,p$;
\item[$Y_i$] is the mass fraction of nucleon  $i$;
\item[$f_a$] is the axion ``decay constant,'' the scale of breaking of the global $U(1)$ symmetry of which the axion is the pseudo-Nambu-Goldstone boson;
\item[$\sigma_{np\pi}$] is the nucleon-nucleon scattering cross section from exchange of a single pion with vanishing pion mass, with canonical value $\sigma_{np\pi} = 4 \alpha_\pi^2 \sqrt{\pi T /m_N^5}$~\cite{Brinkmann:1988vi, Raffelt:1993ix, Raffelt:2006cw}, where 
$\alpha_\pi \simeq 15$ and $T$ is the temperature of the SM matter in the proto-neutron star; 
\item[$\gamma_{\rm f}$] is introduced to cut off the low-energy divergence of \Eq{corrected-rate}~\cite{Keil:1996ju, Sigl:1997ga}. We use the form $\bL 1  + (n_B \sigma_{np\pi}/2\omega)^2 \bR^{-1}$~\cite{Keil:1996ju}, which mimics plasma effects that cut off small-angle scattering;
\item[$\gamma_{\rm p}$] accounts for the finite pion mass and nucleon degeneracy, for which we use the dimensionless phase space integral $s \!\pL n_B, Y_i, \frac\omega T ,\frac{m_\pi}T \pR$ described in the case of neutron-neutron scattering at arbitrary degeneracy in Eq.~(49) of~\cite{Hannestad:1997gc}; and
\item[$\gamma_{\rm h}$] is the ratio of the dynamical spin structure function calculated in chiral perturbation theory for nucleons $i,j$ to the value in the one-pion exchange approximation, schematically $\gamma_{\rm h} = S_\sigma/\!\left.S_\sigma\mR_{\rm OPE}$, for $S_\sigma$ defined in~\cite{Hanhart:2000ae, Bacca:2008yr, Bartl:2014hoa}. The $Y_i=0.5$ case was originally obtained with a nuclear potential calculation by~\cite{Sigl:1997ga}; the $Y_i=0$ case was addressed in~\cite{Hanhart:2000ae} using the soft radiation approximation and measured nucleon phase shifts; and the extension to arbitrary proton fraction using chiral effective field theory at high densities and measured nucleon phase shift at low densities data was developed in~\cite{Bacca:2008yr, Bartl:2014hoa, Bartl:2016iok}. For simplicity, we use the fitting function in Eq.~5 and Tab.~1 of~\cite{Bartl:2016iok} and we assume no energy dependence, which is roughly compatible with~\cite{Sigl:1997ga, Hanhart:2000ae} away from the deuteron resonance at relatively low energies. 
\end{itemize}
We reproduce these various correction factors in \Fig{corrections}, fixing $\omega=T$ for illustration. Critically, each correction factor 
individually reduces the original rate by a non-negligible multiplicative factor. Very roughly speaking, we find that the rates are suppressed by a factor between 5 and 100 from the core to the neutrinosphere. We discuss alternate parameterizations of these effects in \App{alt-corr} and find very similar results. Some of these effects, specifically $\gamma_{\rm f}$, have been included in calculations of the axion luminosity before, 
as discussed in \App{previous-axion}, but this is the first attempt to collect all known effects together. Combined with our improved treatment of the energy dependence of the optical depth and our inclusion of different supernova temperature and density profiles, we find that bounds on the axion mass may change significantly from the canonical values.

Since we will be interested in the sum of the scattering rate over all nucleon pairs, we define a reduced coupling constant $C^2=Y_n C_n^2 +  Y_p C_p^2$.
We then go on to model-independently bound $C^2$ along with the axion decay constant $f_a$. Following convention, we will show this as a bound on the axion mass, which is in one-to-one correspondence with the decay constant. The relation between $f_a$ and $m_a$ is, at leading order, $m_a^2 f_a^2 = m_\pi^2 f_\pi^2 m_u m_d/(m_u+m_d)^2$~\cite{DiVecchia:1980yfw}, and including NLO effects the relation is $m_a= 5.7 \ev (10^6 \gev\!/f_a)$~\cite{diCortona:2015ldu}. Finally, we have
\beq \label{ax-width}
\Gamma_a \simeq 5.2\tenx{-15} \mev \pL \frac{\rho_B}{\rho_c}\pR^2 \pL \frac{T}{T_c}\pR^{1/2} \pL \frac{\omega}{T_c}\pR^{-1}  \frac{C^2}{C^2_{\rm KSVZ}} \pL\frac{f_a}{5.7\tenx6\gev}\pR^{-2} \gamma_{\rm f} \gamma_{\rm p} \gamma_{\rm h}\,,
\eeq
where the density and temperature are of order $\rho_c = 3\tenx{14}\g/\cm^3$, $T_c=30\mev$, and the reduced coupling in the case of the KSVZ axion, with $C_n \simeq 0, C_p\simeq -0.47$~\cite{diCortona:2015ldu}, is $C^2_{\rm KSVZ}\simeq 0.066$ for $Y_p=0.3$. We use \Eq{L-A'} with the replacements $\tau = \int \Gamma_{\rm ibr} dr \to \int \Gamma_a dr $ and $\Gamma_{\rm br} \to e^{-\omega/T} \Gamma_a$ to get the total instantaneous luminosity in axions. The hadronic axion that we consider is stable against decay and other absorptive processes in the proto-neutron star, so \Eq{ax-width} is the only width we need to calculate.

\begin{figure}[t]
\begin{center}
\includegraphics[width=0.49\textwidth]{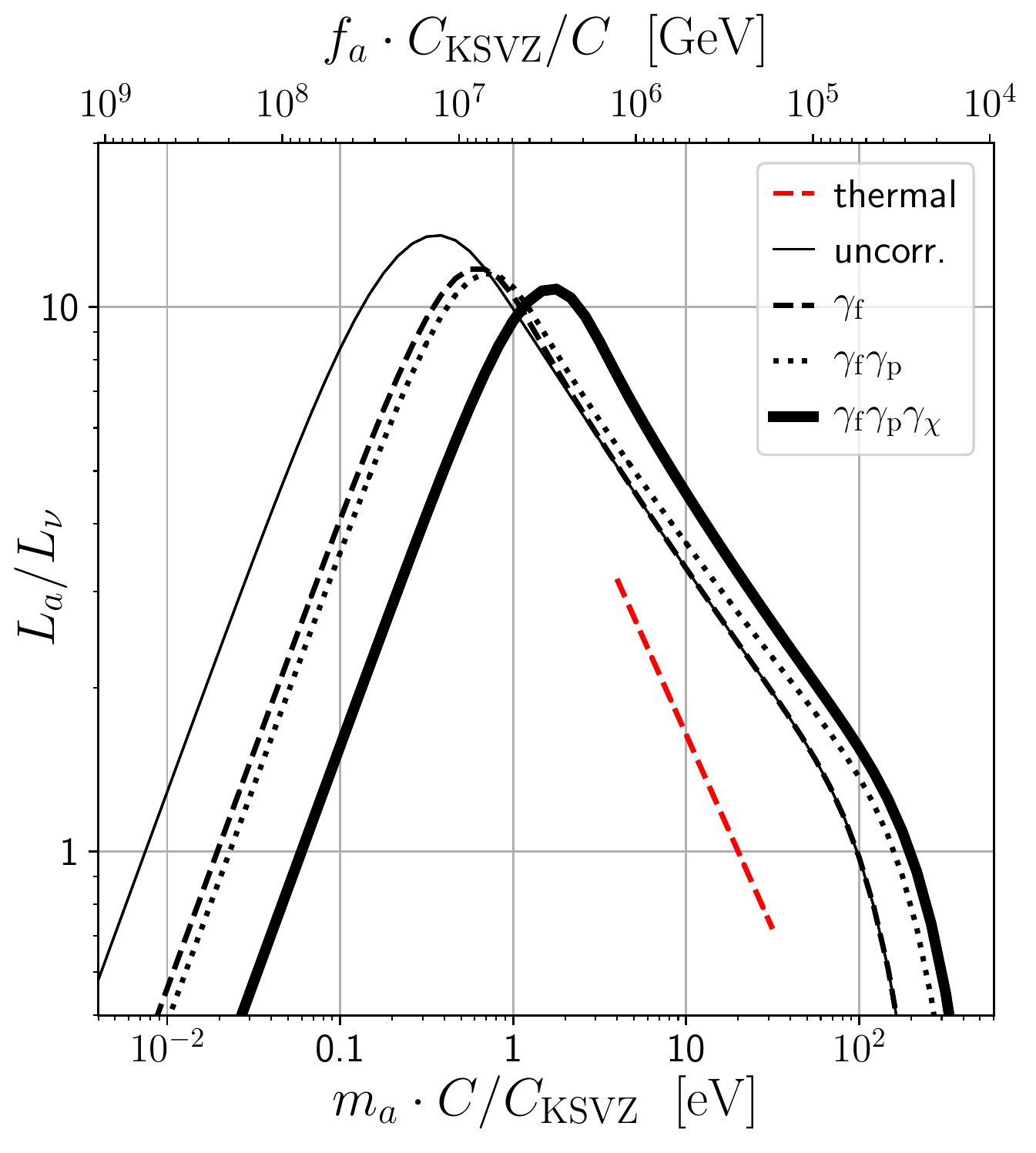}~
\includegraphics[width=0.5\textwidth]{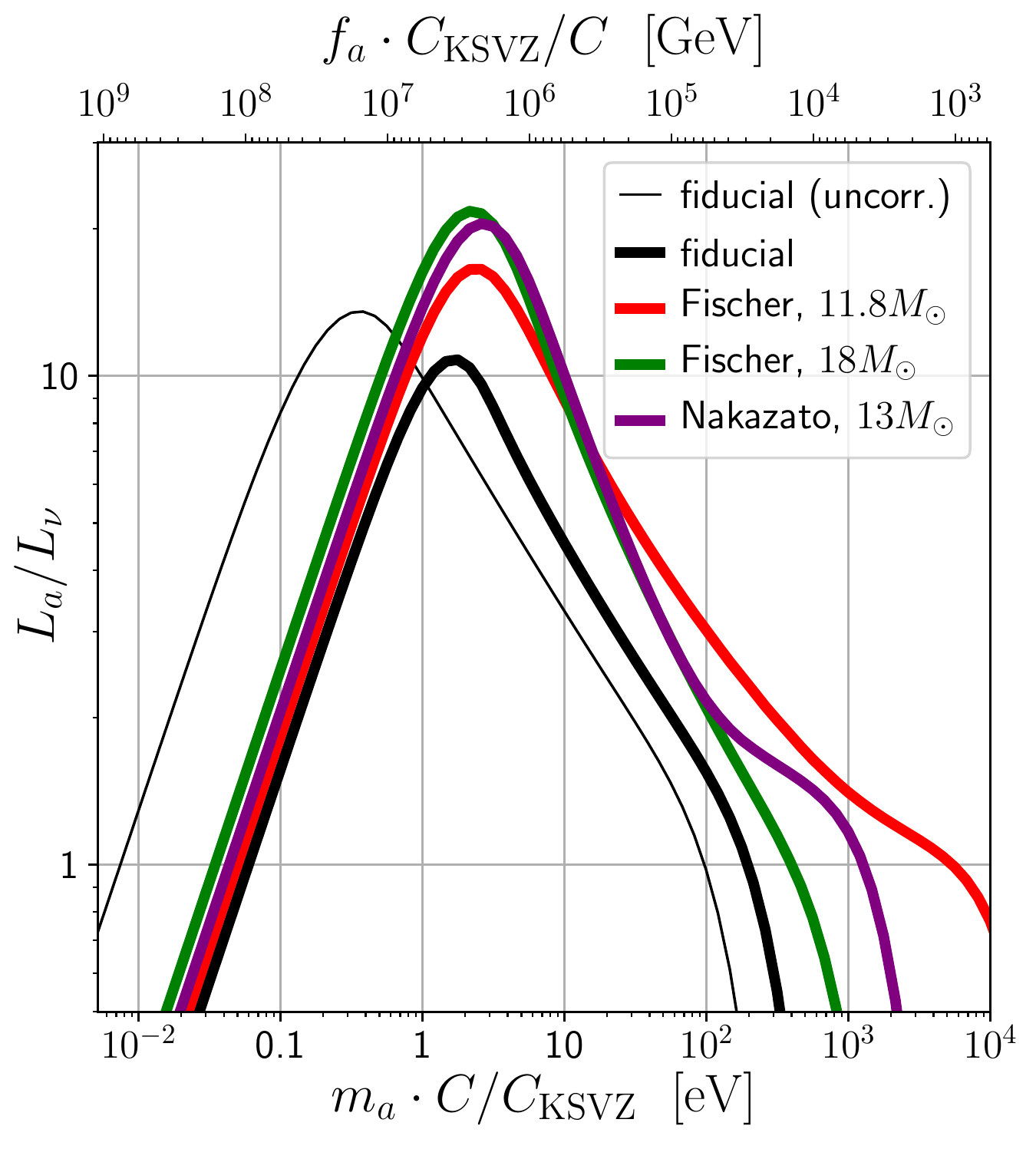}
\caption{
{\bf Left:} 
Luminosity of the QCD axion for a variety of correction factors. 
The red dashed line labelled ``thermal'' is the bound one would obtain at large couplings (equivalently, small $f_a$ or large $m_a$) 
if one assumes that the emission is a blackbody.  
The black lines instead assume a more accurate calculation as described in the text.  
The thin black line labelled ``uncorr.'' does not include any correction factors given in \Eq{corrected-rate}, 
while the other black lines include one, two, or all three correction factors, respectively.   
{\bf Right:}  Luminosity of the QCD axion for a variety of supernova temperature and density profiles. 
}
\label{ax-constraints}
\end{center}
\end{figure}

We emphasize that 
our various correction factors collectively reproduce the N${}^3$LO calculation in chiral perturbation theory~\cite{Bacca:2008yr, Bartl:2014hoa, Bartl:2016iok} and together should consistently ``correct'' the leading order calculation of the axion emission rate. In other words, the product $\gamma_{\rm f} \gamma_{\rm p} \gamma_{\rm h}$ is a self-consistent correction: starting from a simplified calculation for which a closed-form solution is easy to calculate, we wind up with the N${}^3$LO ChPT result. However, a full calculation should include additional effects and error bars. New nuclear potentials could also be used to expand on our treatment of higher-order corrections, \eg~by including additional energy dependence that we did not model. 
It is also important to understand more systematically the exact nature of the low-energy cutoff.  
For these and other reasons, an exact calculation is still desirable.

\subsection{Results}

\begin{figure}[t]
\begin{center}
\includegraphics[width=0.85\textwidth]{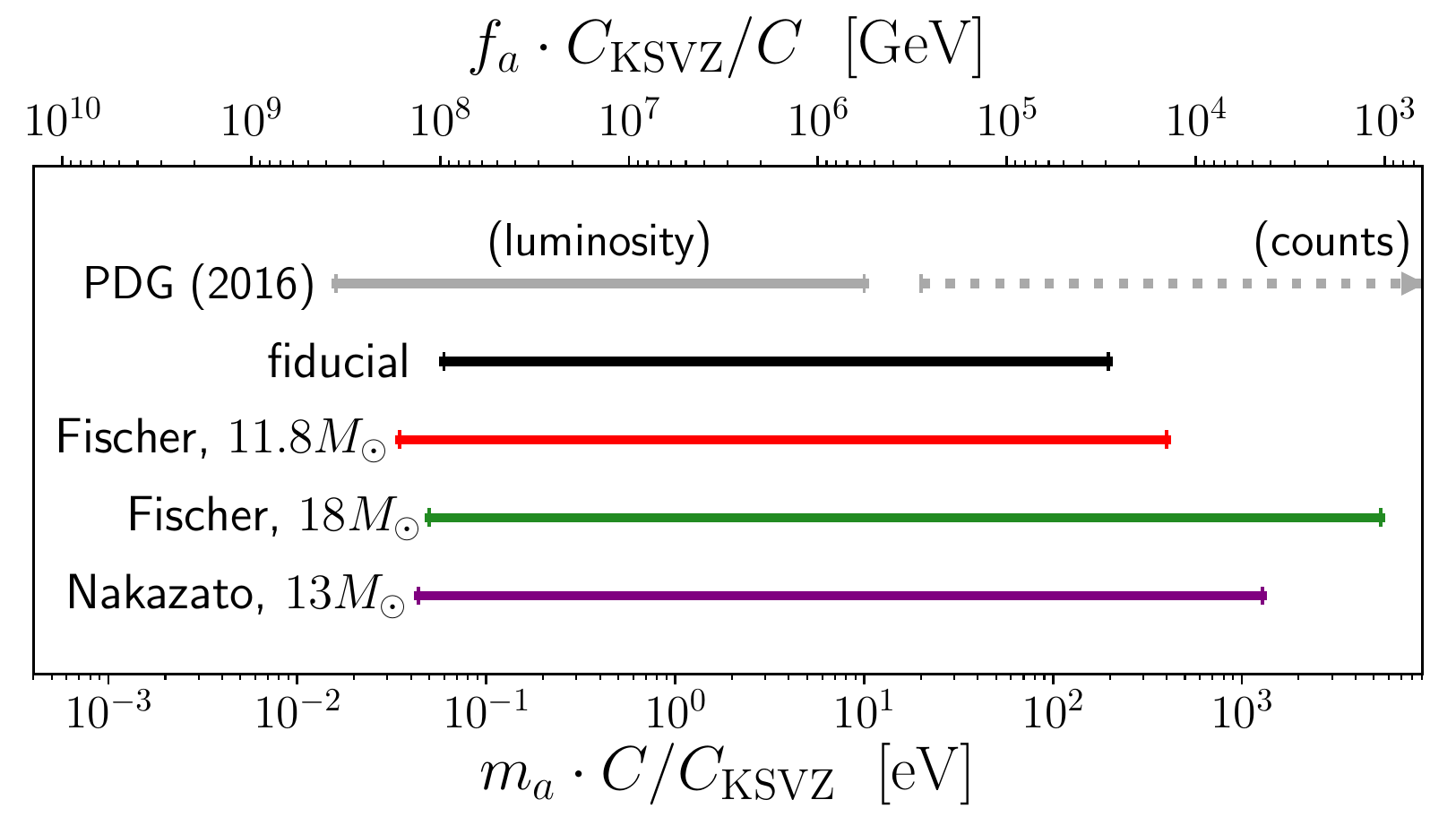}
\caption{Constraints on the QCD axion mass and axion decay constant for various supernova temperature and density profiles.
The ``canonical'' bound from the PDG~\cite{Patrignani:2016xqp, Raffelt:2006cw} is shown with a solid gray line, while the bound labelled ``counts'' 
comes from~\cite{Engel:1990zd}.  Our bounds close the gap between these constraints, known as the ``hadronic axion window''.}
\label{ax-constraints-linear}
\end{center}
\end{figure}

We plot the luminosity\footnote{We emphasize that for the range of couplings where $L_a \gg L_\nu$, backreaction of the axion flux on the star will be qualitatively important for the stellar evolution and the luminosity should not be interpreted literally. For $L_a \sim L_\nu$ (and, in particular, for $L_a = L_\nu$ where we set a bound) the backreaction should be slight and our results should be realistic.} as a function of QCD axion mass times reduced coupling in \Fig{ax-constraints}, and we show the corresponding excluded regions of the axion mass times reduced coupling in \Fig{ax-constraints-linear}. In the left panel of \Fig{ax-constraints}, we show the breakdown of effects arising from the different correction factors $\gamma$ and also from the novel treatment of the optical depth at high coupling. The improvement in the treatment of the optical depth leads to big effects at large coupling, while the low-energy cutoff and higher-order diagrams have bigger effects at low coupling. In the right panel of \Fig{ax-constraints}, and in \Fig{ax-constraints-linear}, we show the effect of using numerical proto-neutron star temperature and density profiles rather than the ``fiducial'' profile adapted from~\cite{Raffelt:1996wa}. Interestingly, we find that the fiducial profile leads to the most conservative excluded region. In all cases, we are able to close the ``hadronic axion window'' that had previously existed between the luminosity bounds~\cite{Raffelt:2006cw} and the bounds from additional counts in Kamiokande for a more tightly coupled axion~\cite{Engel:1990zd}, labeled ``(counts)'' in \Fig{ax-constraints-linear}. We also point out that our revised bound has implications for the claim that stellar cooling anomalies can be explained by weakly coupled, non-hadronic axions \cite{Giannotti:2017hny}, and new joint constraints are warranted.

Our results differ from those in canonical references by up to roughly two orders of magnitude at large coupling and a factor of a few at 
small coupling~\cite{Patrignani:2016xqp, Raffelt:2006cw}. This comes from several effects, all of which point in the same direction. Our approach to taking into account the energy dependence of the optical depth, following~\cite{Chang:2016ntp}, increases the extent of the bounds at high coupling by approximately a factor of five compared to assuming that axions thermalize and are emitted with a blackbody spectrum. The remaining difference between our final bounds and the ones shown in~\cite{Patrignani:2016xqp} is slightly less than an order of magnitude: 
the difference is apparent at both high and low coupling, and is attributable to our inclusion of the correction factors $\gamma$ in \Eq{corrected-rate}. The factors $\gamma_{\rm f}$ and $\gamma_{\rm p}$ lead to approximately a factor of a few discrepancy with~\cite{Patrignani:2016xqp}, and the corrections to the nucleon scattering rate encapsulated by $\gamma_{\rm h}$ lead to a similar correction. We illustrate this breakdown in the left panel of \Fig{ax-constraints}. These nuclear corrections have been incorporated for neutrino interactions in various nuclear physics codes that evolve supernova explosions, in particular in~\cite{Bartl:2016iok}, but to our knowledge this 
is the first time these effects have been included in bounds on the interactions of the axion.

\section{Axion-like Particles with Yukawa Couplings}
\label{alpsection}

\begin{figure}[t]
\begin{center}
\includegraphics[width=0.6\textwidth]{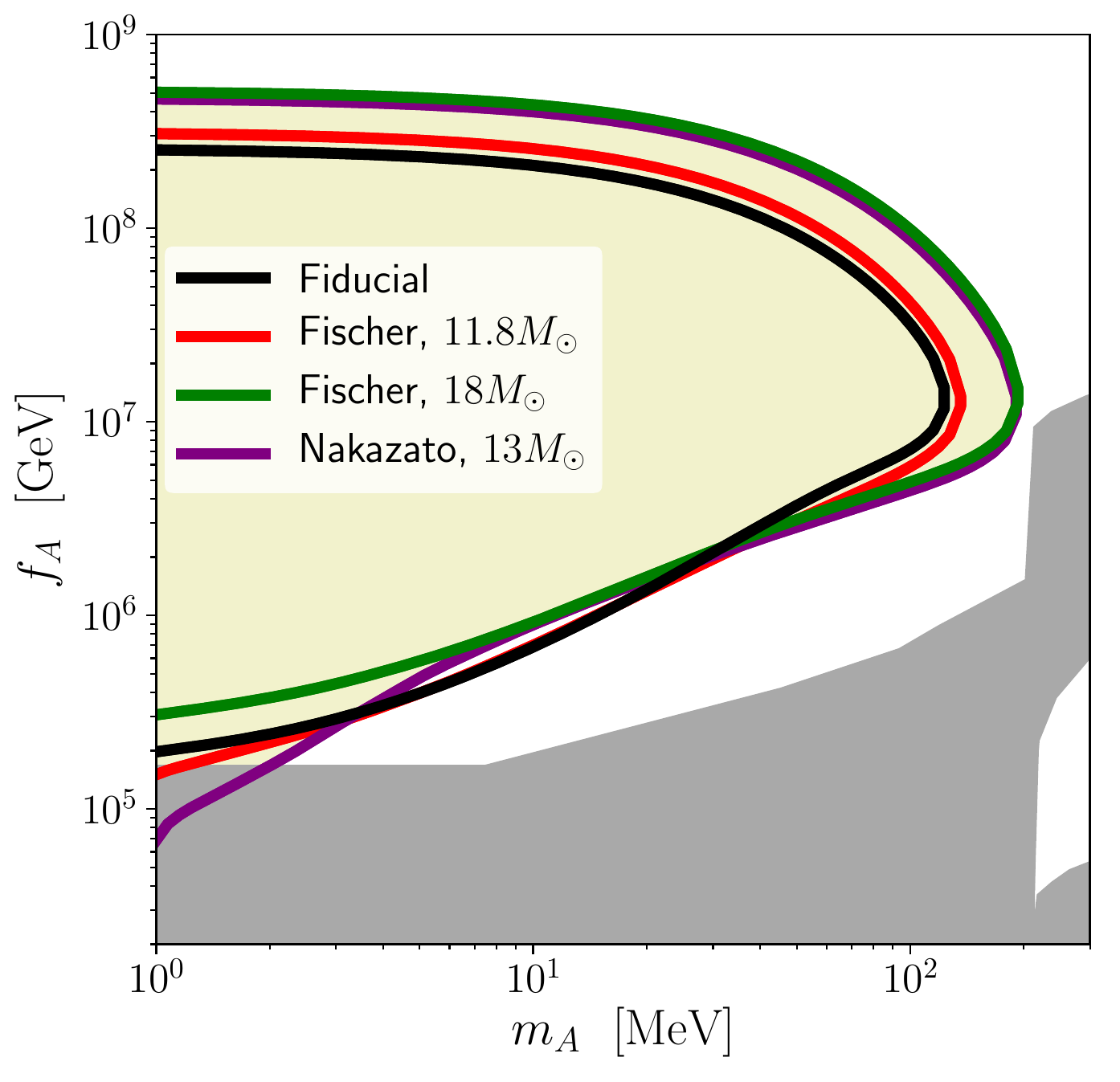}
\caption{Thick solid black, red, green, and red lines show the SN1987A constraints for various temperature and density profiles 
on axion-like particles (ALPs) with Yukawa couplings, updating the bounds presented in~\cite{Essig:2010gu}.  
Other constraints are taken from~\cite{Essig:2010gu,Dolan:2014ska}, omitting bounds due to Kaon decays from \cite{Dolan:2014ska, Kahlhoefer}.
}
\label{alpbound}
\end{center}
\end{figure}

Our analysis of the QCD axion naturally extends to variations on the single-parameter axion model. This allows us to investigate bounds in a general two-parameter space for what is commonly referred to as an ``axion-like particle,'' or ALP. The ALP mass $m_A$ and ``decay constant'' $f_A$ are not related, and we will explore the part of the parameter space for which the finite mass of the ALP becomes non-negligible 
(our bounds can then be simply extrapolated to lower masses at fixed $f_A$).

As in \cite{Essig:2010gu}, we consider an axion-like particle for which the mass and coupling are no longer related, $m_A f_A \slashed\simeq \Lambda_{\rm QCD}^2$. The Lagrangian for the ALP scenario is similar to the QCD axion Lagrangian, but the ALP couples to particles other than the nucleons, 
\bea \cL \supset \sum_i \frac{C_i}{2 f_A} \partial_\mu a \bar f_i \gamma^\mu \gamma_5 f_i\,. 
\eea
Since the mass and coupling are now independent parameters, we assume for simplicity that all of the couplings $C_i$ are equal, but because of the identity $\partial_\mu \bar f \gamma^\mu \gamma_5 f \to 2i m_f \bar f \gamma_5 f$ noted above, the ALP will couple less strongly to lighter SM fermions. For this reason, as well as for the reasons enumerated at the beginning of \Sec{subsec:LA'}, the ALP production due to scattering or annihilation of electrons in the proto-neutron star is negligible. The luminosity is then given by
\alg{
&L_A = \int_0^{R_\nu} dV \int \frac{d^3 k_a}{(2\pi)^3} \omega e^{-\omega/T} \Gamma_A \times
\\ & \qquad \times \exp\cbL - \int_0^{R_{\rm far}} dr \bL \Gamma_A + \frac{m_A}{8\pi} \sum_\ell  \Theta(m_A-2m_\ell)\frac{m_\ell^2}{f_A^2} \sqrt{1-\frac{4m_\ell^2}{m_A^2}} \bR \cbR,
}
where we define $\Gamma_A = \Gamma_a\times \sqrt{1-m_A^2/\omega^2},$ with $\Gamma_a$ from \Eq{ax-width}. For convenience, we have 
rescaled $f_A \to f_A/C_i$. We explicitly include a contribution to the absorptive width of the axion for its decay to leptons $\ell$ if $m_a>2m_\ell$, although in practice we find that this does not affect the limits at all, since $m_e \ll T$ and the Boltzmann suppression effectively depresses the production rate for $m_a \gtrsim 2m_\mu$. 
We assume that the coupling to leptons does not affect the early stages of the supernova explosion, but this must be checked for self-consistency. If we omit these couplings the SN1987A bounds are not impacted, but 
the accelerator and rare-decay bounds from \cite{Essig:2010gu,Dolan:2014ska} are not applicable.

We show our results in \Fig{alpbound}, updating the bounds in~\cite{Essig:2010gu}. In particular, as in the DM case, axions 
with masses that are kinematically accessible but too weakly coupled to be produced at accelerator-based experiments are potentially probed by SN1987A~\cite{Essig:2010gu}. 
Interestingly, a small gap remains between the SN1987A bounds and accelerator-based searches.

\section{Conclusion}
\label{conclusionsection}

In this paper, we have considered constraints derived from the duration of the neutrino cooling phase of SN1987A on two broad 
classes of DM particles: a dark sector fermion coupled to a kinetically mixed dark photon, as well as the QCD axion and axion-like particles.

For the dark sector fermion, we derive constraints for several different mass hierarchies of the dark-sector particles.  
We show these constraints for the case of a heavy dark photon that can decay to dark fermions ($m' = 3m_\chi$) for elastic DM 
in \Figs{lightdmplots}{lightdmconstplots} and for inelastic DM in \Fig{idmplots}; for 
the case of a massive but light dark photon ($m'\lesssim m_\chi \lesssim T_c$) in \Figs{heavydmplots1}{heavydmplots2}; 
and for the millicharged case ($m' \ll m_\chi \lesssim T_c$) in \Fig{millichargeddmplots}.  
To derive these constraints, we have suggested a novel criterion for highly mixed dark fermions, wherein they return to chemical equilibrium if they take a random walk in their velocity vector that turns them $90^\circ$ from their initial direction of motion. We use this requirement because the scattering cross section for dark sector fermions can be very forward peaked, and scattering an $\mathcal{O}(1)$ number of times does not change a light DM trajectory enough to prevent the DM from escaping. 

Our bounds have important implications for popular sub-GeV DM models, in which the DM couples to a dark photon of similar mass.  
They suggest that large regions of otherwise unexplored sub-GeV DM parameter space are now disfavored.  
However, the SN1987A bounds are complementary to both existing bounds and proposed experimental searches: 
they lie well below current bounds, and many motivated and concrete benchmark-model 
``targets'' remain unconstrained.  This further emphasizes the need for a robust experimental program to search 
for sub-GeV DM as envisioned in~\cite{Battaglieri:2017aum}, at least down to the SN1987A constraint, if not beyond.  

The QCD axion has been studied in some detail previously, but bounds on axion properties from SN1987A have heretofore been extracted with a range of simplifying assumptions that are known to be violated at the order-of-magnitude level. Here we attempted to rectify this situation by including some estimates of known nuclear physics and particle physics effects. In particular, recent progress in chiral effective theories demonstrates that corrections up to N${}^3$LO can have a substantial impact on the spin fluctuation rate of free nucleons~\cite{Bacca:2008yr, Bartl:2014hoa, Bartl:2016iok}, confirming earlier calculations using nuclear phase shift data~\cite{Sigl:1997ga, Hanhart:2000ae} that have long been applied to the neutrino emissivity. These effects conspire to point in the same direction, resulting in large changes to the expected axion emission rate. Coupled with our improved description of boson luminosity in the high-mixing limit, the axion bounds are changed 
significantly 
from the ``canonical'' range, as shown in \Fig{ax-constraints-linear}.
We also re-visited the constraints on axion-like particles with Yukawa couplings, shown in \Fig{alpbound}, 
finding some difference with the previous literature. 

The wealth of information that has been gained over the years from the observation of SN1987A is rather remarkable.  
As simulations of core-collapse supernova keep improving, it will be highly desirable to continue the effort to include 
new, weakly-coupled particles directly into the simulations.

\section*{Acknowledgements}
We thank Alex Bartl, Savas Dimopoulos, Vera Gluscevic, Peter Graham, Roni Harnik, Eder Izaguirre, Felix Kahlhoefer, Gordan Krnjaic, Ken'ichiro Nakazato, Gustavo Marques Tavares, 
Surjeet Rajendran, Harikrishnan Ramani, Annika Reinert, Philip Schuster, Achim Schwenk, Jordan Smolinsky, Natalia Toro, Michael Turner, and William Wester for useful discussions and/or correspondence. 
RE~and JHC~acknowledge support from DoE Grant DE-SC0017938.  SDM acknowledges support from the YITP when this work commenced. For the second half of this work, SDM was supported by Fermi Research Alliance, LLC under Contract No. DE-AC02-07CH11359 with the U.S. Department of Energy, Office of Science, Office of High Energy Physics. The United States Government retains and the publisher, by accepting the article for publication, acknowledges that the United States Government retains a non-exclusive, paid-up, irrevocable, world-wide license to publish or reproduce the published form of this manuscript, or allow others to do so, for United States Government purposes.

\begin{appendix}

\section{Production and Decay of Dark Photons}
\label{aprime-prod}
In the relativistic, degenerate regime we use Eq.~(77) of~\cite{Braaten:1993jw} with the conventions of~\cite{An:2013yfc} to define the SM photon polarization tensor:
\beq  \label{pol-ten}
\begin{array}{ll}
\re \Pi_L = \frac{3\omega_p^2}{v^2}  \pL 1-v^2 \pR \bL \frac1{2v} \ln \pL \frac{1+v}{1-v}  \pR -1 \bR, &\qquad\omega_p^2 = 4\pi \alpha_{\rm EM} n_e/E_F
\\  \re \Pi_T =\frac{3\omega_p^2}{2v^2} \bL 1 - \frac{1-v^2}{2 v} \ln \pL \frac{1+v}{1-v}  \pR \bR,  &\qquad E_F^2 = m_e^2 + (3\pi^2 n_e)^{2/3}.
\end{array}
\eeq
As a result of the structure of the mixing angle, there is a particular energy $\omega_*$ at which $\re\Pi = m'^2$, where the mixing angle hits a resonance. When production is resonant, the magnitude of the differential power exactly compensates the narrow width of the resonance, 
and the luminosity becomes independent of the production mechanism. The rates for resonant inverse bremsstrahlung and electromagnetic decay of the $A'$ particle are~\cite{Chang:2016ntp}
\alg{ \label{wid-abs}
\Gamma_{\rm ibr.}^{L,T} &= \frac{32}{3\pi} \frac{\alpha_{\rm EM} (\epm)_{L,T}^2 n_nn_p}{\omega^3}\pL \frac{\pi T}{m_N}\pR^{3/2}  \langle \sigma_{np}^{(2)}(T) \rangle \bL \frac{m'^2}{\omega^2} \bR_L
\\ \Gamma_e^{L,T} &= \frac{\alpha_{\rm EM} (\epm)_{L,T}^2 m'^2}{\sqrt{\omega^2-m'^2}} \int_{x_-^e}^{x_+^e} d x \frac1{\exp\pL\frac{-x+\mu_e/\omega}{T/\omega}\pR+1} \times \cbL \begin{array}{cc} m_e^2/m'^2+ z(x) & ~(T) \\
1 - 2z(x) & ~(L) \end{array} \cbR
}
where $\langle \sigma_{np}^{(2)}(T) \rangle =\frac12  \int_0^\infty dx \int_{-1}^1 d \cos \theta\,e^{-x} x^2 \frac{d\sigma_{np}(x)}{d\theta}$ is an angle- and energy-averaged neutron-proton scattering cross section extracted from measured nuclear phase shifts~\cite{Rrapaj:2015wgs}; we introduce a kinematic function $z(x) =x \pL \frac{\omega}{m'}-x \pR - \pL \frac12 - \frac{\omega x}{m'} \pR\bL \frac12 -\frac{\omega}{m'}\pL \frac{\omega}{m'}-x \pR  \bR \frac1{\omega^2/m'^2-1}$; and the endpoints of the energy integral are $x_\pm^e = \frac12 \pm \frac12 \sqrt{(1- 4m_e^2/m'^2)\pL1-m'^2/\omega^2\pR }$ for the outgoing electron-positron pair. In the soft radiation approximation, the detailed balance factor $e^{-\omega/T}$ between bremsstrahlung and inverse bremsstrahlung becomes unity, and so we define a bremsstrahlung production rate 
\beq
\Gamma_{\rm br.}^{L,T} = \frac{32}{3\pi} \frac{\alpha_{\rm EM} (\epm)_{L,T}^2 n_nn_p}{\omega^3}\pL \frac{\pi T}{m_N}\pR^{3/2}  \langle \sigma_{np}^{(2)}(\omega,T) \rangle \bL \frac{m'^2}{\omega^2} \bR_L,
\eeq
where $\langle \sigma_{np}^{(2)}(\omega, T) \rangle =\frac12 \int_{\omega/T}^\infty dx  \int_{-1}^1 d \cos \theta \,e^{-x} x^2 \frac{d\sigma_{np}(x)}{d\theta}$ differs from $\langle \sigma_{np}^{(2)}(T) \rangle$ only in the lower limit of the energy integral.

\section{Dark Matter Bremsstrahlung}
\label{chi-prod}

One of the dominant production modes for DM in the supernova is via on- or off-shell $A'$ bremsstrahlung during nucleon elastic scattering events, as in \Fig{feyn-diags}. We calculate this amplitude of this process in the limit of soft bremsstrahlung following \S6.1 of~\cite{Peskin:1995ev}. 

The matrix element for DM production is
\alg{ \label{mat-el-0}
i\cM&=ie\bar{u}(P_3) \bL \cM_{np}(P_3,P_1-K)\frac{i(\slashed P_1-\slashed K +m_N) \gamma^\mu}{(P_1-K)^2-m_N^2}  + \frac{i\gamma^\mu (\slashed P_3+\slashed K +m_N)}{(P_3+k)^2-m_N^2} \cM_{np}(P_3+k,P_1) \bR u(P_1) \times \\
& \times \left (\frac{\cP_{L \mu \nu}}{K^2-\Pi_L}+\frac{\cP_{T \mu \nu}}{K^2-\Pi_T} \right ) i \epsilon K^2 g^{\nu \alpha} \frac{i(-g_{\alpha \beta}+K_\alpha K_\beta /m'^2)}{K^2-m'^2+i m' \Gamma_\chi} \bar{u}(\chi)i g_D \gamma^\beta v(\bar{\chi}),
}
where $\cM_{np}$ is the matrix element for the process with no bremsstrahlung, which is $n-p$ scattering~\cite{Rrapaj:2015wgs}; the incoming $p$ [$n$] has four momentum $P_1^\mu = (E_1, \vec p_1)$ [$ P_2^\mu = (E_2,\vec p_2)$]; the outgoing $p$ [$n$] has four momentum $P_3^\mu =(E_3,\vec p_3)$ [$P_4^\mu=(E_4,\vec p_4)$]; the DM particles have four momenta $P_\chi^\mu = (E_\chi,\vec p_\chi)$ and $P_{\bar \chi}^\mu=(E_{\bar \chi} ,\vec p_{\bar \chi})$; and the dark photon carries an interior momentum $K^\mu = P_\chi^\mu + P_{\bar \chi}^\mu = (\omega, \vec k)$. In what follows, lower-case letters without the vector symbol indicate the magnitude of the three vector, {\it e.g.} $k = |\vec k|$. We include different contributions from the longitudinal and transverse modes, which can contribute differently in the dense environment of the proto-neutron star.

In the low momentum or ``soft'' limit, \Eq{mat-el-0} becomes
\alg{ \label{mat-el-1}
\cM&=\epsilon e\bar{u}(P_3) \cM_{np}(P_3,P_1) u(P_1) \left ( \frac{2P_1^\mu}{K^2-2P_1 \cdot K} + \frac{2P_3^\mu}{K^2+2P_3 \cdot K} \right)  \times \\ & \qquad \qquad \qquad \times \left (\frac{\cP_{L \mu \nu}}{K^2-\Pi_L}+\frac{\cP_{T \mu \nu}}{K^2-\Pi_T} \right )  \frac{K^2}{K^2-m'^2+i m' \Gamma_\chi} \bar{u}(\chi)g_D \gamma^\nu v(\bar\chi)\,.
}
We square the amplitude and take the trace. Current conservation, $K^\mu \cP_{\mu\nu}=0$, implies $ \cP_{\mu \nu}(P_\chi^\nu P_{\bar{\chi}}^\beta+P_{\bar{\chi}}^\nu P_\chi^\beta)\cP_{\alpha \beta} = -2\cP_{\mu\nu} P_\chi^\nu P_\chi^\beta \cP_{\alpha \beta}$, leading to
\alg{
|\cM|^2&=-32 \pi^2 \epsilon^2 \alpha \alpha_D |\cM|_{np}^2 \frac{K^4}{(K^2-m'^2)^2+(m'\Gamma_\chi)^2}  \times  \\ &\times  \left ( \frac{2P_1^\mu}{K^2-2P_1 \cdot K} + \frac{2P_3^\mu}{K^2+2P_3 \cdot K} \right) \left ( \frac{2P_1^\alpha}{K^2-2P_1 \cdot K} + \frac{2P_3^\alpha}{K^2+2P_3 \cdot K} \right) \times
\\ &\times \left (\frac{\cP_{L \mu \nu}}{K^2-\Pi_L}+\frac{\cP_{T \mu \nu}}{K^2-\Pi_T} \right )\left (\frac{\cP_{L \alpha \beta}}{K^2-\Pi_L^*}+\frac{\cP_{T \alpha \beta}}{K^2-\Pi_T^*} \right )
(4P_\chi^\nu P_\chi^\beta+K^2 g^{\nu \beta})\,.
}
We find that we may in general rearrange this as
\alg{
|\cM|^2 &= \frac{ |\cM|_{np}^2 | \vec p_1 - \vec p_3|^2}{m_N^2 } \cS( K, P_\chi)\,,
}
where, for the sake of brevity, we separate the contribution due to the $n-p$ dynamics from a ``soft factor'' $\cS$ due to the DM emission. This soft factor is a function only of the virtual and radiated particle momenta.

Assuming that the DM does not scatter on its way out of the star, we can calculate the local differential luminosity per unit volume,
\alg{
\frac{dL_\chi}{dV}= \int\frac{d^3 \vec p_1\,f_1}{(2\pi)^3 2E_1} \frac{d^3\vec p_2\,f_2}{(2\pi)^3 2E_2} \frac{d^3\vec p_3\, (1-f_3)}{(2\pi)^3 2E_3} \frac{d^3\vec p_4\, (1-f_4)}{(2\pi)^3 2E_4} \frac{d^3\vec p_\chi \, (1-f_\chi)}{(2\pi)^3 2E_\chi}  \frac{d^3\vec p_{\bar \chi} \, (1-f_{\bar \chi})}{(2\pi)^3 2E_{\bar \chi}} \times \\ \times (2\pi)^4\delta^4(P_1+P_2-P_3-P_4-P_\chi-P_{\bar\chi})  \frac{ |\cM|_{np}^2 \mL\vec p_1 - \vec p_3\mR^2}{m_N^2 } \omega \cS( K, P_\chi)\,,
}
where the $f_i$ are distribution functions. In the following, we will assume the particles are non-degenerate such that we may ignore all $(1-f)$ factors. We also approximate the effect of the soft radiation limit (invoked above to obtain the matrix element) by taking $\delta^4(P_1+P_2-P_3-P_4-P_\chi-P_{\bar\chi}) \simeq \delta^4(P_1+P_2-P_3-P_4)e^{-\omega/T}$. Note we use a different approximation from~\cite{Chang:2016ntp}, which gives more conservative results. Changing variables $P_{\bar\chi} \to K$ for convenience, we have
\alg{ \label{dPdV-1}
\frac{dL_\chi}{dV} &= \int\frac{d^3\vec p_1 f_1}{(2\pi)^3 2E_1} \frac{d^3\vec p_2 f_2}{(2\pi)^3 2E_2} \frac{d^3\vec p_3}{(2\pi)^3 2E_3} \frac{d^3\vec p_4}{(2\pi)^3 2E_4} (2\pi)^4\delta^4(P_1+P_2-P_3-P_4) \frac{  |\cM|_{np}^2 |\vec{p_1}-\vec{p_3}|^2}{m_N^2}\times \\
 & \qquad \times\int\frac{d^3\vec p_\chi}{(2\pi)^3 2 E_\chi} \int\frac{d^3\vec k}{(2\pi)^3 2(\omega-E_\chi)} \omega e^{-\omega/T} \cS( K, P_\chi) \,.
}
Since the first line only depends on the nucleon scattering and the second line is only sensitive to the DM kinematics, we can calculate them separately.

The first line of \Eq{dPdV-1} involves many of the same features as the result in~\cite{Chang:2016ntp} and summarized in \App{aprime-prod}, and we follow a similar procedure. In particular, we assume the nucleons are nonrelativistic and invoke the relations
\beq
|\cM|_{np}^2=64\pi^2E^2_{\cm}\frac{d\sigma_{np}}{d\Omega_{\cm}}, ~~~ f_{1,2}=n_{p,n} \left (\frac{2\pi}{m_N T} \right )^{3/2} e^{-|\vec{p}_i|^2/2MT}, ~~~ T_{\rm CM} = \frac{(\vec p_1 - \vec p_2)^2}{4m_N}
\eeq
to get
\beq
\text{\Eq{dPdV-1}, first line} \approx \frac{16}{\sqrt{\pi}}\left( \frac{T}{m_N} \right)^{3/2} n_n n_p \langle \sigma_{np}^{(2)} (T) \rangle,
\eeq
where $\langle \sigma_{np}^{(2)}(T) \rangle $ is defined below \Eq{wid-abs}. With a little work, the second line of \Eq{dPdV-1} is
\alg{
\int&\frac{d^3\vec p_\chi}{(2\pi)^3 2 E_\chi} \int\frac{d^3\vec k}{(2\pi)^3 2(\omega-E_\chi)} \omega e^{-\omega/T} \cS( K, P_\chi) =  \frac{ 256 \pi^4 \alpha_{\rm EM} \alpha_D \ep^2}3 \times \\ \times & \int d|\vec p_\chi|d|\vec k| d \cos \theta_{k \chi} |\vec p_\chi|^2 |\vec k|^2 \omega e^{-\omega/T} \cbL \frac{k^4\bL k^4 - 4\pL E_\chi |\vec k| - \omega |\vec p_\chi| \cos \theta_{k \chi}\pR^2 \bR}{\omega^4 \bL \pL k^2 - m'^2 \pR^2 - (m' \Gamma_\chi)^2 \bR \bL \pL k^2 - \re \Pi_L \pR^2 - \im \Pi_L^2 \bR} \right. +
\\ & \qquad \left.+ \frac{2 k^4\pL k^2 - 2|\vec p_\chi|^2 \sin \theta_{k \chi}^2 \pR}{\omega^2 \bL \pL k^2 - m'^2 \pR^2 - (m' \Gamma_\chi)^2 \bR \bL \pL k^2 - \re \Pi_T \pR^2 - \im \Pi_T^2 \bR }+ {\text{L-T cross-terms}} \cbR,
}
which is calculable numerically.

\section{Dark Matter Elastic Scattering}
\label{DMscat}

Once DM particles are produced, they elastically scatter off protons on their way out of the supernova, as in \Fig{feyn-diags}. This can lead to the thermalization of the DM particles, which can allow them to return to chemical equilibrium, as described in \Sec{subsec:trapping}. The matrix element squared for this process is
\alg{ \label{DMscat-Msq}
|\cM|_s^2&=16 \pi^2 \ep^2 \alpha \alpha_D \frac{K^4}{(K^2-m'^2)^2+(m'\Gamma_\chi)^2}  \left (\frac{\cP_{L \mu \nu}}{K^2-\Pi_L}+\frac{\cP_{T \mu \nu}}{K^2-\Pi_T} \right )  \left (\frac{\cP_{L \alpha \beta}}{K^2-\Pi_L^*}+\frac{\cP_{T \alpha \beta}}{K^2-\Pi_T^*} \right ) \times\\ 
&\times \tr[\gamma^\mu(\slashed{P}_1+m_\chi)\gamma^\alpha(\slashed{P}_3+m_\chi)]\tr[\gamma^\nu(\slashed{P}_2+m_N)\gamma^\beta(\slashed{P}_4+m_N)]\,,
}
where $P_1$($P_2$), $P_3$($P_4$) are incoming and outgoing DM (proton) momenta and $K=P_1-P_3$ is the momentum transfer. We define a scattering rate and an average angular deflection per scatter by
\begin{align}
\Gamma_s&=\frac{1}{2E_1} \int \frac{d^3\vec p_2\,f_2}{(2\pi)^3 2E_2} \frac{d^3\vec p_3}{(2\pi)^3 2E_3} \frac{d^3\vec p_4}{(2\pi)^3 2E_4} (2\pi)^4\delta^4(P_1+P_2-P_3-P_4) |\mathcal{M}|_s^2
\\ \Delta \theta &= \frac{1}{2E_1\Gamma_s} \int \frac{d^3\vec p_2\,f_2}{(2\pi)^3 2E_2} \frac{d^3\vec p_3}{(2\pi)^3 2E_3} \frac{d^3\vec p_4}{(2\pi)^3 2E_4} (2\pi)^4\delta^4(P_1+P_2-P_3-P_4) \theta_{13} |\mathcal{M}|_s^2\,,
\end{align}
where $\theta_{13}$ is an angle between incoming and outgoing DM, and we assume the protons and the DM are nondegenerate, such that $1-f_3 \simeq 1-f_4 \simeq 1$. Also, we assume $f_2$ follows the Maxwell-Boltzmann distribution. With these definitions, we can define the total number of scatters and the maximum angular deflection in \Eq{expected-deflection}. 

Scattering through a light mediator has a $t$-channel singularity and thus is weighted towards small angles, $\theta_{13}|_{\rm typical} \ll \pi/\few$. This characteristically ``forward peaked'' scattering indicates that light DM scattering through a light mediator neither loses a significant fraction of its energy nor deviates far from its initial trajectory in a typical scattering event. Thus, measures of DM decoupling that assume stationary initial state nucleons and calculate mixing angles for which the DM scatters an order one number of times are bound to overestimate the tendency of DM to be trapped in the neutron star and thus to predict an overly small value of the kinetic mixing as the trapping line.

We also point out that if the dark photon is massless, as we assume in \Sec{millichargesection}, then the longitudinal mode decouples and the propagators $\cP_L$ should be omitted from \Eq{DMscat-Msq} because the different polarizations do not mix.

Finally, note that in \Eq{expected-deflection}, we numerically evaluate the integrals from $R_d$ to $R_\nu$, 
and we approximate the integrand for the region from $R_\nu$ and $R_f$ as $\Gamma(R_\nu) R_\nu/5 v$. 

\section{Alternate Parameterizations of Axion Corrections}
\label{alt-corr}

In \Sec{axion-corrections} we chose to correct the rate for axion bremsstrahlung to match the energy-averaged rate calculated at N${}^3$LO order in chiral perturbation theory. Other parameterizations of similar effects exist in the literature, and the exact rate could in principle deviate from the chiral perturbation theory result. Here, we summarize some possible alternative correction factors that account for similar physical effects in different ways:
\begin{itemize}
\item[$\gamma_{m_\pi}$] accounts for the finite pion mass rather than $\gamma_{\rm p}$, which we model as $\gamma_{m_\pi} = \pL 1 + \frac{m_\pi^2}{3m_NT} \pR^{-2}$, roughly matching~\cite{Raffelt:1993ix} (this analytic  prescription falls between the numerical work of~\cite{Stoica:2009zh}, obtained with non-degenerate nucleons, and the result of~\cite{Iwamoto:1984ir, Iwamoto:1992jp}, calculated for degenerate nucleons);
\item[$\gamma_{\rm SRA}$] is the ratio of the dynamical spin structure function for nucleons $i,j$ in the soft radiation approximation to the value in the one-pion exchange approximation with finite pion mass, obtained numerically with the aid of nuclear phase shift measurements. The $Y_i=0.5$ case was originally discussed by~\cite{Sigl:1997ga} while the $Y_i=0$ case was addressed in~\cite{Hanhart:2000ae}, each finding reductions of order a few. For the purpose of illustration, we will neglect the density dependence and simply assume a constant factor of 5 reduction in the rate compared to the uncorrected result, which potentially underestimates the axion luminosity; and
\item[$\gamma_{\rm LPM}$] accounts for the Landau-Pomeranchuk-Migdal effect, for which we use the semi-analytic fit $\gamma_{\rm LPM}= \bL 1 + \frac13\pL \frac{\rho}{\rho_c} \pR^{1/3} \bR^{-6}$ following~\cite{Fischer:2016boc}.
\end{itemize}
Together, $\gamma_{\rm SRA}$ and $\gamma_{\rm LPM}$ should roughly combine to account for the same physics as $\gamma_{\rm f}$ and $\gamma_{\rm h}$ in our main results. Likewise, the factor $\gamma_{m_\pi}$ potentially mimics the effect of the pion propagator in place of $\gamma_{\rm p}$.

\begin{figure}[t]
\begin{center}
\includegraphics[width=0.46\textwidth]{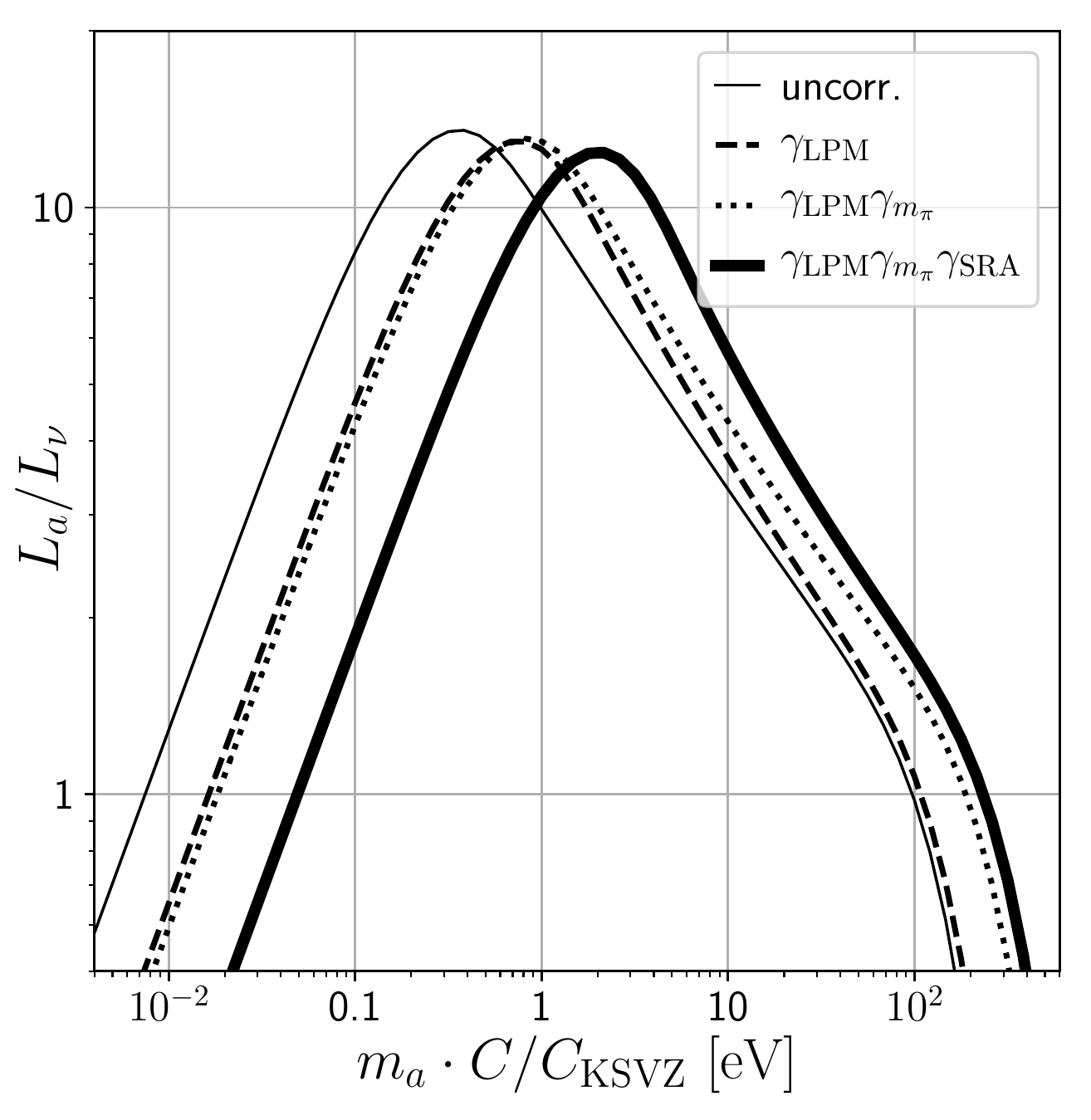}~~
\includegraphics[width=0.47\textwidth]{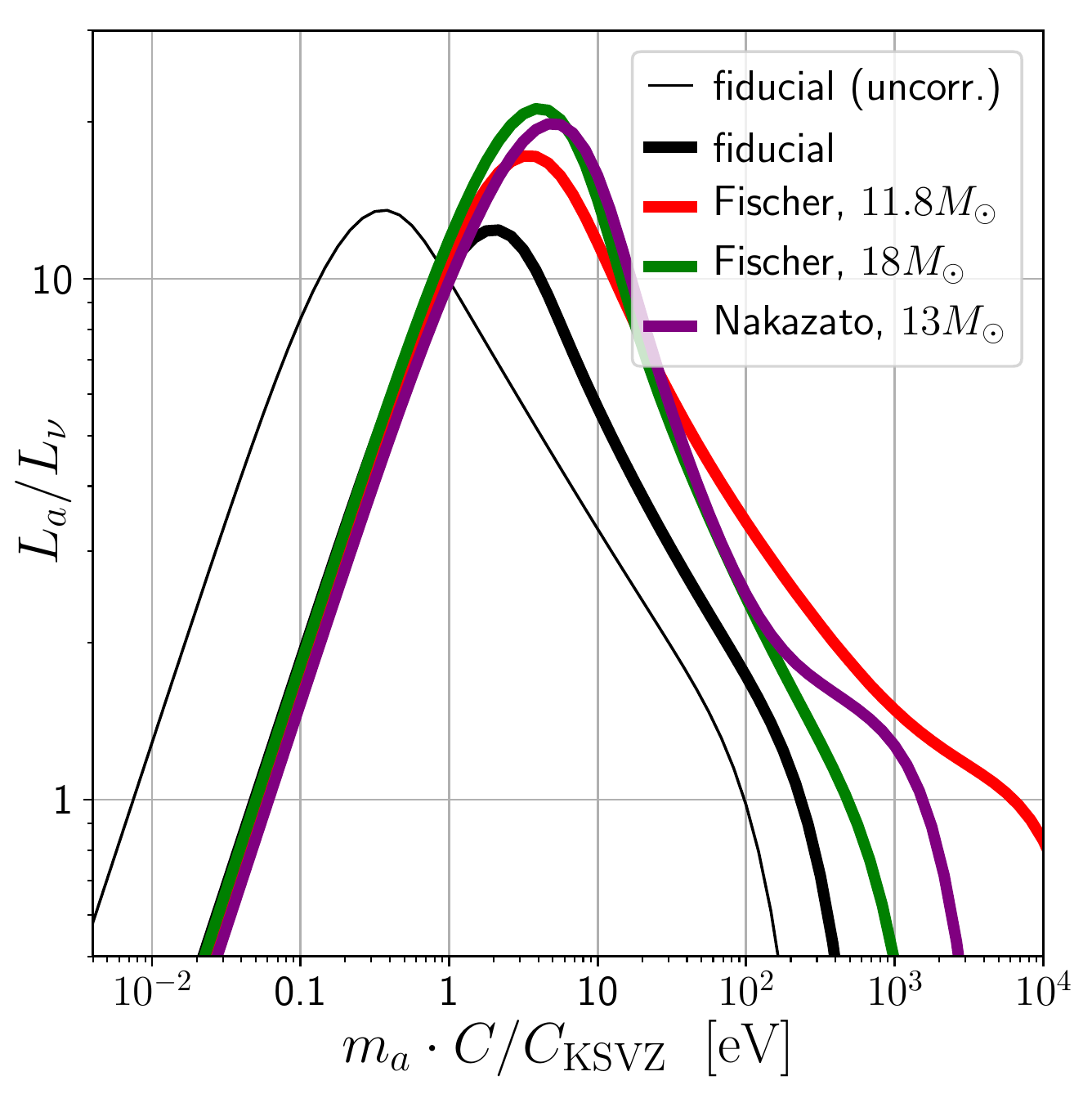}
\caption{Luminosity of the QCD axion for a variety of alternate correction factors and supernova profiles, as described in \App{alt-corr}. 
This figure is similar to \Fig{ax-constraints}, and by comparing these two figures, we find that the excluded regions are very similar. 
}
\label{ax-constraints-alt}
\end{center}
\end{figure}

We show results for the luminosity in axions using these alternate correction factors in \Fig{ax-constraints-alt}. This figure is the counterpart of \Fig{ax-constraints}. By comparison of these two figures we see that the excluded regions change very little regardless of how we choose to account for these nuclear physics corrections.

\section{Summary of Previous Work on the Hadronic Axion}
\label{previous-axion}

The absorption rate of the QCD axion has been obtained to varying degrees of precision since before the explosion of SN1987A. Because the axion is predominantly produced during nucleon-nucleon bremsstrahlung, the exact result requires evaluating the fifteen-dimensional integral of a non-perturbative matrix element with a partially degenerate phase space. This technical challenge has taken quite some time to thoroughly understand.

Here we summarize the evolution of the work that has previously put bounds on the QCD axion, listed in chronological order:
\begin{itemize}
\item[\cite{Turner:1987by, Raffelt:1987yt}] provided the first calculations for axions emitted from nucleon-nucleon bremsstrahlung, modeling the nuclear interaction with a single (massless) pion exchange and assuming that the squared matrix element is constant in the nucleon momenta;
\item[\cite{Turner:1988bt}] compared measured pion production rates in $p-p$ scattering to those found from a diagrammatic one-pion exchange calculation and found that these agreed to within a factor of a few;
\item[\cite{Burrows:1988ah}] conducted supernova explosion simulations including a free-streaming axion energy sink and backreaction on the star for a wide variety of proto-neutron star profiles;
\item[\cite{Brinkmann:1988vi}] computed phase space integrals over the one-pion exchange diagram for arbitrary nucleon degeneracies, justifying the use of non-degenerate phase space;
\item[\cite{Burrows:1990pk}] conducted supernova explosion simulations for a tightly coupled axion, confirming prior bounds in the trapping regime;
\item[\cite{Raffelt:1993ix}] verified the calculation of~\cite{Brinkmann:1988vi} and discusses ways of cutting off pathological limits, including the first appearance of the $1/(\omega^2 + a\Gamma^2)$ prescription and a discussion of when the pion mass should not be neglected;
\item[\cite{Keil:1996ju}] also advocates the $1/(\omega^2 + a\Gamma^2)$ approach and additionally proposes a ``saturation width'' that cuts off the rates at some maximum spin fluctuation rate.
\end{itemize}
All of these authors roughly agree in the free-streaming limit, finding the requirement that the Peccei-Quinn scale must respect $f_a \gtrsim 10^{8-9} \gev$, with the uncertainty on this limit primarily arising from the difference in treatment of the low-energy scattering, which can be cut off by the $1/(\omega^2 + a\Gamma^2)$ factor.

Many other works have calculated neutrino couplings and luminosities, which are important because neutrinos and axions couple to the same nuclear current. There are too many developments to name here, but we do clarify the origin of the chiral effective theory corrections that we utilize above:
\begin{itemize}
\item[\cite{Sigl:1997ga}] gives the spin density structure function for $n-p$ scattering in a variety of ways, indicating a qualitative difference in the magnitude of the scattering rate and in the density dependence, ultimately due to the different contribution to the partition function of $n-p$ scattering, which can be resonant near the formation of a deuteron;
\item[\cite{Hanhart:2000ae}] gives the ratio of spin density structure function for identical nucleon scattering based on measured phase shifts;
\item[\cite{Bacca:2008yr, Bacca:2011qd, Bartl:2014hoa, Bartl:2016iok}] use a chiral effective theory approach at high densities and show that this matches to the phase shift analyses at intermediate densities, all of which confirm the high-density suppression and low-density enhancement suggested by~\cite{Sigl:1997ga, Hanhart:2000ae}
\end{itemize}
These corrections are a major ingredient that lead us to the modified limits shown in \Fig{ax-constraints}.

\end{appendix}

\bibliography{DM-SNbib}
\bibliographystyle{jhep}

\end{document}